\documentclass[12pt]{article}
\usepackage{amsmath}
\usepackage{graphicx,psfrag,epsf}
\usepackage{enumerate}
\usepackage{natbib}
\usepackage{url} 

\newcommand{\blind}{1}

\usepackage{bm}
\usepackage{amssymb}
\usepackage{caption, subcaption}
\usepackage{setspace}
\newtheorem{theorem}{Theorem}

\newcommand{\bbeta}{\bm{\beta}}
\newcommand{\bx}{\bm{x}}
\newcommand{\bY}{\bm{Y}}

\newcommand{\btheta}{\bm{\theta}}

\graphicspath{{figures/}}

\begin{document}

\def\spacingset#1{\renewcommand{\baselinestretch}%
{#1}\small\normalsize} \spacingset{1}


\if1\blind
{
  \title{\bf Segmented correspondence curve regression model for quantifying reproducibility of high-throughput experiments}
  \author{Feipeng Zhang   \hspace{.2cm}\\
  Department of Statistics, Hunan Normal University, Hunan, 410081, China\\
  Frank Shen  \hspace{.2cm}\\
    Department of Statistics, Pennsylvania State University,  PA, 16802, USA\\
    Tao Yang \hspace{.2cm} \\
    Bioinformatics and Genomics Program, Pennsylvania State University,  \\PA, 16802, USA\\
    and\\ 
    Qunhua Li 
    \thanks{
    Corresponding author. Email: qunhua.li@psu.edu.
    The authors gratefully acknowledge the financial support of  
    NIH R01GM109453 and 
    the National Natural Science Foundation of China (11771133). 
   }
   \\   
    Department of Statistics, Pennsylvania State University,  PA, 16802, USA
   }
  \maketitle
} \fi

\if0\blind
{
  \bigskip
  \bigskip
  \bigskip
  \begin{center}
    {\LARGE\bf Segmented correspondence curve\\
     regression model for quantifying reproducibility of high-throughput\\ experiments}
\end{center}
  \medskip
} \fi

\bigskip
\begin{abstract}
The reliability of a high-throughput biological experiment relies highly on the settings of the operational factors in its experimental and data-analytic procedures. Understanding how operational factors influence the reproducibility of the experimental outcome is critical for constructing robust workflows and obtaining reliable results. One challenge in this area is that candidates at different levels of significance may respond to the operational factors differently.
To model this heterogeneity, we develop a novel segmented regression model, 
based on the rank concordance between candidates from different replicate samples,
to characterize the varying effects of operational factors for candidates at different levels of significance.  A grid search method is developed to  identify the change point in response to the operational factors and estimate the covariate effects accounting for the change. 
A sup-likelihood-ratio-type test is proposed to test the existence of a change point. 
Simulation studies show that our method yields a well-calibrated type I error, 
is powerful in detecting the difference in reproducibility, and achieves a better model fitting than the existing method. 
An application on a ChIP-seq dataset reveals interesting insights on how sequencing depth affects the reproducibility of experimental results, demonstrating the usefulness of our method in designing cost-effective and reliable high-throughput workflows.

\vspace*{10pt}
\noindent%
\end{abstract}

\vfill

\newpage
\spacingset{1.45} 

\section{Introduction.}\label{sec:intro}

High-throughput technologies are indispensible tools in modern biological research. In each high-throughput experiment, thousands of candidates are measured simultaneously, and the ones that show significant relevance to the biological feature of interest are identified for downstream analyses.
The reliability of the output from a high-throughput experiment largely depends on the operational factors in the experimental and data analytical procedures, such as sequencing depth and the amount of starting materials in a sequencing experiment. 
How to choose parameters for these factors is a critical question in the design of high-throughput experiments. 
An important step towards making proper design choices is to understand how these factors affect the reproducibility of the experiments. Reproducibility between replicated experiments, while not an assessment of the accuracy of a workflow, is indicative of its fidelity and reliability. It plays an important role in establishing confidence of a high-throughput workflow, since the ground truth is usually unavailable for a full evaluation of its sensitivity and specificity.
  

The final output of a high-throughput experiment is often represented in the form of a ranked list, 
where the candidates measured in the experiment are ordered by the significance (e.g. p-values) of their association with the biological feature of interest. Table \ref{tab:ChIP}a shows an example of five workflows, each with  two replicate experiments.
How concordantly the candidates are ranked in replicate experiments is a common criterion to assess the reproducibility of a workflow. 
To properly assess a workflow, a researcher much account for several characteristics of high-throughput experiments
 \citep{li2011measuring}. 
First, 
the level of reproducibility varies for candidates at different levels of prominence. Strong candidates generally are more reproducible than weak ones. 
This is because the strong candidates, which are usually truly biologically relevant, are more likely to be replicated in replicate experiments. 
Second, while it is desirable to have a good concordance for all candidates, the concordance of strong candidates usually are more of a concern, since they are the likely targets of follow-up analyses.  
Third, 
the reproducibility of findings from a high-throughput experiment depends on the stringency of the threshold used to identify the findings. For the same data, the findings based on a stringent threshold usually would have a higher concordance (e.g.  a higher rank correlation) than those based on a relaxed threshold. 
Therefore, the reproducibility of findings at a given threshold is not enough to reflect the reproducibility of the workflow, though the former is often used as a proxy for the latter.
Instead, a proper evaluation of the reproducibility of a workflow should focus on the intrinsic property of the workflow, rather than being tied to a specific threshold. 


Recently, several statistical methods that account for these characteristics have been proposed for assessing and comparing the reproducibility of high-throughput workflows. 
The correspondence curve \citep{li2011measuring} and the correspondence at the top (CAT) plot \citep{irizarry2005multiple} are graphical tools that display how the concordance of entries on two rank lists changes with their ranks.
They can be used to visualize the reproducibility of a workflow between a pair of replicated experiments. 
The reproducibility of workflows with different operational parameters can be compared by comparing their curves.  A quantitative alternative is the irreproducible discovery rate (IDR) approach \citep{li2011measuring}.  This method evaluates the probability that a candidate is a reproducible discovery 
using a copula mixture model, and reports the expected rate of irreproducible discoveries in the selected candidates in a fashion analogous to the false discovery rate. 
The reproducibility of different workflows can be compared by comparing the  numbers of reproducible candidates across replicates passing a series of IDR levels.
Maximum Rank Reproducibility (MaRR), a non-parametric alternative to the IDR method based on a maximum rank statistics, is also available \citep{philtron2017maximum}. 
These methods have been used to compare reproducibility of different workflows in ChIP-seq \citep{li2011measuring}, microarray \citep{irizarry2005multiple}, and RNA-seq studies \citep{philtron2017maximum}. 
However, none of them quantifies the difference in reproducibility between workflows or infers the statistical significance of the difference.
They also have difficulty identifying the individual effect of a factor when multiple factors are studied simultaneously. 

More recently, \citet{li2017regression} proposed a regression model, called the correspondence curve regression, to assess the effect of operational factors on the reproducibility of high-throughput workflows. 
By modelling the correspondence curve using a cumulative regression model, it allows one to quantify the simultaneous and independent effects of operational factors on reproducibility without being linked to a specific significance threshold.
However, this method implicitly assumes that reproducibility changes with significance at a constant rate. 
This assumption is likely to be violated when strong and weak candidates have considerably different reproducibility or different responses to operational factors. 
For example, in a ChIP-seq experiment, a high-throughput experiment for identifying protein binding sites on the DNA, the sites with strong binding affinities
usually are identified more reproducibly than those with weak binding affinities \citep{landt2012chip}.
These two types of sites also respond differently to the increase of sequencing depth, an important operational factor in ChIP-seq experiments.
The strong sites usually show marked improvement in reproducibility at a low sequencing depth and little improvement after reaching a sufficient sequencing depth, 
whereas the weak sites 
require a much higher sequencing depth to show improvement 
  (Figure \ref{fig:ChIP}). This suggests that sequencing depth has obviously different covariate effects on reproducibility between these two types of binding sites. 
 Dissecting the effects of sequencing depth between these two types of sites will provide more insights on how to choose the optimal sequencing depth that achieve sufficient reproducibility -- a critical question in the design of ChIP-seq experiments.
 More details can be found in Section \ref{S:motivation}. 

In this article, we develop a segmented correspondence curve regression model with an unknown change point to address this issue.
 The key idea of this method is to view the difference in reproducibility between strong and weak candidates as a structural change in association along with the significance level. 
 This naturally allows us to embed the correspondence curve regression into a segmented regression framework, such that the effects of these two groups and the transition point are characterized in a seamless framework.
  The segmented regression framework 
  is a useful tool for studying the changes of the relationship between response and covariates. It typically comprises of two or more segments with different slopes that intersect at change points, and is suitable for modeling data with a continuous segmented relation. 
Many variants have been proposed in different contexts, for example, 
least squares regression 
\citep{quandt1958estimation, hinkley1969inference, feder1975asymptotic, chappell1989fitting, 
chiu2006bent, hansen2017regression},
quantile regression \citep{lich2011bent}, 
expectile regression \citep{zhang2017threshold}, 
and rank-based regression \citep{zhang2017robust}. 
However, these models mostly focus on the change in the mean structure, rather than the change in association. 

To estimate the parameters, we propose a grid-search method to simultaneously estimate the change point and the covariate effects on reproducibility for candidates on both sides of the change point.  
Using the modern empirical processes theory, we derive the asymptotic properties for all the parameters including the change point, and show that the change-point estimator achieves root-n consistency. 
Furthermore, we propose a formal test procedure to test the existence of a change point in reproducibility, based on the sup-likelihood-ratio-type test statistic in \cite{lee2011testing}.   
Although there are many tests for the existence of a change point in different contexts,
to our knowledge, no test has been developed for the existence of a transition of reproducibility.

Altogether, our methods allow one to more accurately describe the influence of the operational factors for both strong and weak candidates, 
providing investigators with a comprehensive picture to 
choose proper experimental parameters. 
More broadly, the proposed methods can be used to detect and quantify changes in concordance in any area where understanding consistency between repeated measurements or comovement between correlated variables is important.  
The method is implemented in the publicly available R package \textit{segCCR} \citep{zhang2018segCCR}. 

The rest of the paper is organized as follows. Section \ref{S:motivation} describes a critical question in the design of ChIP-seq experiments that motivates this work: how many reads should be sequenced to obtain reliable results in a cost-effective manner? 
In Section~\ref{sec:method}, 
we introduce the main methodology, including an estimation procedure for the the segmented correspondence curve regression, its asymptotic properties, and a test for the existence of a change point. 
In Section~\ref{sec:numerical},  we demonstrate  our method using simulations.  
In Section~\ref{sec:applications}, we apply our method to a ChIP-seq study in \citet{chen2012systematic} to determine the cost-effective sequencing depth that achieves reasonable reproducibility. We also demonstrate the use of our method as a general tool for estimating the structural change of association using a salmon dataset.   
Section~\ref{sec:conclusion} concludes the paper with discussions. 
All the technical proofs are presented in Supplementary.

\section{Motivating Applications}\label{S:motivation}

{
One important consideration in designing ChIP-seq experiments is to determine the desired sequencing depth, i.e. the number of reads to sequence. 
Sequencing depth is known to directly affects the sensitivity and specificity of the detection of binding sites from a ChIP-seq experiment \citep{landt2012chip, chen2012systematic}.  
If sequencing depth is insufficient, binding sites may not be detected reproducibly.
Increasing sequencing depth can improve the detection of binding sites. 
It has been observed that weaker biologically functional sites, such as low-affinity sites or regions of open chromatin that bind transcription factors less specifically, can be identified more reliably at a deeper sequencing depth \citep{landt2012chip}.   
Inevitably, higher sequencing depth would incur higher costs.  
Despite the rapid reduction of sequencing costs over the past decade, it is still expensive to generate deeply sequenced data, especially if many samples need to be sequenced. 
Interestingly, previous studies also reported that increasing reads beyond a certain minimum number has diminishing return. 
For example, 
when using a deeply sequenced ChIP-seq data to identify the binding sites of a point-source transcription factor, Su(HW), in Drosophila, \citet{chen2012systematic} found that more than half of the Su(HW) peaks detected at 120 million reads were captured at 2.7 million reads.  Thus it is of great interest to determine the minimum sufficient sequencing depth that achieves robust results.

The sufficient sequencing depth depends on
the purpose of the downstream study \citep{landt2012chip, jung2014impact}, 
 especially on the importance of weak sites to the downstream analysis.
For binding motif discovery, a relatively low depth that allows the identification of strong binding sites would be sufficient. A high depth is required to obtain a comprehensive collection of all candidate regulatory regions bound by a transcription factor.
Though there are some general guidelines for the minimum sequencing depth in a ChIP-seq experiment \citet{landt2012chip, jung2014impact}, 
it is still unclear how deep one should sequence for a specific experimental purpose.
A detailed characterization of how sequencing depth affects the reproducibility of strong and weak sites will provide investigators a clear understanding on the impact to downstream studies. 

Here we consider a deeply sequenced high-quality ChIP-seq data set in \citep{chen2012systematic}. 
This dataset was generated from Drosophila S2 cells for the site-specific transcription factor Suppressor of Hairy-wing (Su(Hw)). 
It consists of 120 million (M) reads, with a depth of ~1 read per base pair (bp) of mappable fly genome, corresponding to ~2.4 billion reads in the human genome.  
This deeply sequenced dataset provides an ideal benchmark for studying the impact of sequencing depth on ChIP-seq experiments.   
Using this dataset, \citet{chen2012systematic} systematically evaluated how  sequencing depth and other experimental parameters influence the identification of binding regions from ChIP-seq experiments. 
To evaluate the impact of sequencing depth, they randomly sampled reads from the complete dataset (120M) to obtain the subsamples with 0.45 M, 0.9 M, 2.7 M, 5.4 M and 16.2 M reads, and identified the binding regions in each subsample using several  peak calling algorithms. 
The detailed experimental design can be found in \citet{chen2012systematic}.  Figure \ref{F:chipseq} shows the ChIP-seq profiles at the five sequencing depths for a region on chromosome 6. 

\begin{figure}[t]
	\centering
		\includegraphics[height=0.4\textheight, width=1.03\textwidth]{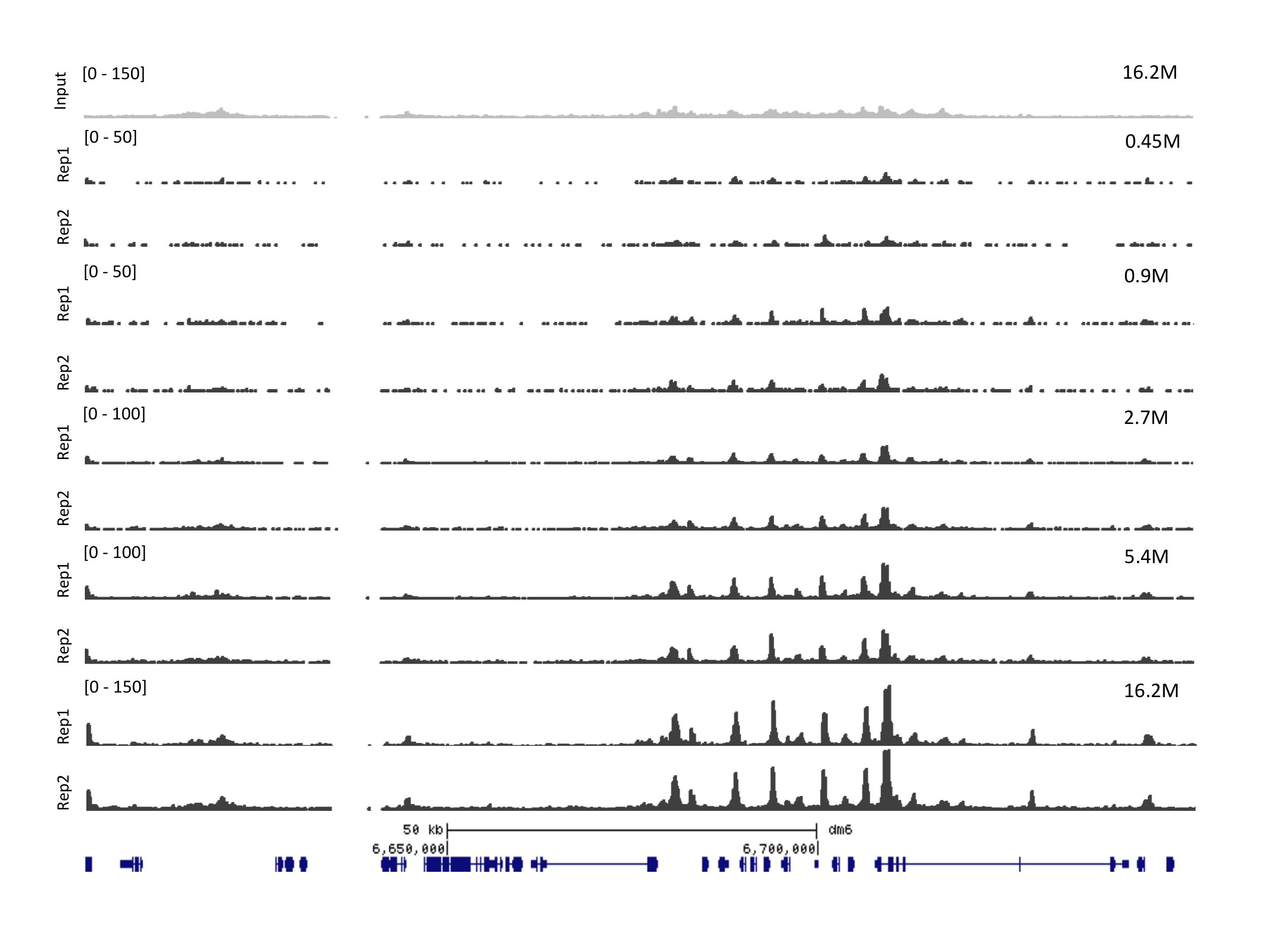}
		\caption{ChIP-seq reads profiles for fly Su(HW) at different sequencing depths in a pair of replicate samples for a gene-rich region on chromosome 6. Numbers on the y-axis denote the ranges of scale. An input profile (i.e. a control without ChIP signals) is on the top row for comparison.}\label{F:chipseq}
\end{figure}		

At each sequencing depth, they assessed the reproducibility of each algorithm across a pair of replicate subsamples using the copula mixture model in \citet{li2011measuring} and 
reported the number of reproducible binding sites at the irreproducible discovery rate of 0.05. 
 For most algorithms, the number of reproducible sites initially increased with sequencing depth and then showed signs of saturation after the sequencing depth reached a certain level.  
This indicates that it is feasible to use reproducibility as a criterion to determine the minimum sequencing depth that achieves reliable identification. 
However, comparing the number of reproducible sites is not the optimal approach for two main reasons.
First, the number of reproducible sites depends on the number of identified binding sites. Even if two datasets have equally reproducible signals, the one with more identified binding sites would have more reproducibly detected sites than the one with fewer identified binding sites. Second, 
the number of reproducible sites does not quantify the statistical significance of the difference in reproducibility, making it difficult to objectively determine when sufficient reproducibility is achieved.    
We believe that a regression-based characterization 
of the covariate effect of the sequencing depth to the reproducibility of strong and weak binding sites
will provide a more direct and comprehensive understanding of 
how sequencing depth affects the reliability of the ChIP-seq experiments, 
helping investigators determine the cost-effective sequencing depth 
for their desired downstream analyses.   
}

\section{Methodology.}\label{sec:method}
The data consist of the outputs of high-throughput experiments 
generated from $S$ workflows ($S\geq 2$). 
All the workflows measure the same underlying biological process,  
but they may differ in certain operational factors, such as sequencing depth or the amount of cells to perform the experiments. For each workflow, few replicate experiments are available such that the reproducibility of the workflow can be assessed across replicates. 
In each replicate experiment, a large number of candidates are evaluated and each candidate is assigned by the workflow a continuous score that reflects the strength of evidence for a candidate to be a true signal. 
The scores can be either original measurements (e.g., fold enrichment) or test statistics obtained based on the original measurements (e.g., p-value). 
We assume all scores are observed and will use scores as the data for our model.  Table \ref{tab:ChIP} shows the data matrix for the example in Section \ref{sec:applications}, which has five workflows with two replicate samples each. 
As in the previous studies \citep{li2011measuring, li2017regression, philtron2017maximum},  
we focus on the case of two replicates here. 

Let $\bY_1^s,\cdots,\bY_n^s$, where $\bY_i^s\equiv(Y_{i, 1}^s, Y_{i, 2}^s)$,  be the significance scores  of a sample of $n$  candidates on two replicates assigned by the workflow $s$. We assume that $Y_{1,j}^s,\cdots, Y_{n,j}^s$  are independent and identically distributed samples from an unknown distribution $F_j^s$ 
for $j=1, 2$. 
Without loss of generality, we assume that low scores represent strong evidence of being genuine signals. 
Given a cutoff $t \in (0,1)$, a candidate is deemed as significant on replicate $j$ if its score is among top $100t\%$, i.e.
$Y^s_{i,j} \leq {\mathbb{F}_j^s}^{-1}(t)$, where $\mathbb{F}_j^s$ is the empirical distribution of $F_j^s$ and ${\mathbb{F}_j^s}^{-1}$ is its inverse.
A candidate is deemed as a reproducible identification from the workflow $s$, 
if its scores are significant on both replicates, i.e. $Y_{i,1}^s\leq {\mathbb{F}_1^s}^{-1}(t)$ and $Y_{i,2}^s\leq {\mathbb{F}_2^s}^{-1}(t)$. For scoring systems that represent strong evidence using high values,
 one just reverses the order correspondingly. 
For notational simplicity, we omit the superscript $s$ when no confusion arises thereinafter.

\subsection{Review of the correspondence curve regression (CCR)}\label{SS:CCR}

To motivate our methods, we first give a brief introduction to the correspondence curve and the correspondence curve regression.

The correspondence curve \citep{li2011measuring} is a graphical tool for visualizing how the concordance between two rank lists changes with the significance of the candidates. Given a set of presepecified thresholds, $\mathcal{T}=\{t \mid 0< t_1 <\ldots<t_M \leq 1\}$  ($M\leq n$),
the correspondence curve sequentially evaluates the proportion of common entries that pass the threshold $t$ among all candidates,
\[
\Psi_n(t) = n^{-1}\sum_{i=1}^{n} \left(Y_{i,1}\leq \mathbb{F}_1^{-1}(t), 
Y_{i,2}\leq \mathbb{F}_2^{-1}(t)\right),
\]
and plots the pairs of $(t, \Psi_n(t))$ for all $t \in\mathcal{T}$. 
The strength of concordance between the rank lists and how the reproducibility changes with the significance of candidates can be visualized.
When applied to the ranked candidate lists of two replicate experiments from a workflow, the correspondence curve reflects the reproducibility of the workflow without being linked to any threshold. 

The correspondence curve regression (CCR) is a parametric regression model that evaluates how covariates affect the reproducibility of the high-throughput experiments \citep{li2017regression}.
The key idea of CCR is that it views $\Psi_n(t)$ as a sequential dichotomization of the scores $(Y_{i1}, Y_{i2})$ at a series of significance thresholds, and then models its relationship with operational factors and thresholds using a cumulative regression framework.
This framework provides a natural way to evaluate
 the covariate effects without being linked to a specific threshold, similar to the correspondence curve.
Specifically, when only one workflow is considered, the baseline model is
\begin{equation}\label{mod-baseline}
g(\Psi(t)) = \sum_{k=1}^K \alpha_k h_k(t), \quad t \in \mathcal{T},
\end{equation}
where 
$g$ is a known link function, $\bm{h}=(h_1, \ldots, h_K)$ are prespecified functions, and $\bm{\alpha}=(\alpha_1, \ldots, \alpha_K)$ is an unknown parameter vector that reflects how the probability to be reproducibly identified changes with the probability of being identified on a single replicate through $\bm{h}$. 
The model that incorporates the operational factors as covariates is
\begin{equation}\label{E:glm}
g(\Psi(t\mid \bm{x})) = \sum_{k=1}^K \alpha_k h_k(t)+ \bm{W}(t, \bm{\beta})^T \bm{x}, \quad t \in \mathcal{T},
\end{equation}
where 
$\bm{W}(t, \bm{\beta})^T \bm{x}$ is a linear predictor characterizing the effect of the covariates $\bm{x}$ on reproducibility, 
and $\bm{\beta}= (\beta_{11},\ldots, \beta_{1K},\ldots, \beta_{d1},\ldots, \beta_{dK})^T$ are unknown coefficients to be estimated. 
Here  $\bm{W}(t, \bm{\beta})=\left(W_1(t, \bm{\beta}_1), \ldots, W_d(t, \bm{\beta}_d)\right)^T$, 
with $W_p(t, \bm{\beta}_p) = \sum_{k=1}^K \beta_{pk} h_k(t)$, where $(h_1, \ldots, h_K)$ is the same set of functions as the one for the baseline terms, and $\beta_{pk}$ measures the covariate effect on $h_k(t)$ due to $x_p$ for $p=1,\ldots, d$. 

The functional form of $g$ and $h_k$ and the value of $K$ can be determined based on a connection between this model and the Archimedean copula families \citep{nelsen2006introduction} established in \citet{li2017regression}. 
That is, for a class of Archimedean copulas, their diagonal sections can be algebraically represented as (\ref{mod-baseline}) with a specific form of $g$ and $h_k$, which is referred to as the canonical functional form. 
 This connection suggests a principled way to select the functional form of $g$ and $h_k$, analogous to the selection of the link function in a generalized linear model: 
one can first find the suitable Archimedean copula model according to the empirical distribution of $(Y_1, Y_2)$, and then specify $g$ and $h_k$ according to its canonical functional form.
For example,  the dependence structure of scores from replicate high-throughput experiments usually follows the shape of the Gumbel-Hougaard copula (Figure~\ref{fig1}d). This implies that the regression form corresponding to the Gumbel-Houggard copula,
\begin{equation}\label{mod1}
g(\Psi(t)) = \alpha h(t), 
\quad t \in \mathcal{T}, 
\end{equation} 
where $g(\cdot)=h(\cdot)=\log(\cdot)$, would provide a reasonable fit of the data. 
A list of Archimedean copulas that satisfy this relationship and their corresponding $g$ and $h_k$ functions are provided in \citet{li2017regression}.




\subsection{Segmented correspondence curve regression model} 



\begin{figure}
	\centering
 	\begin{subfigure}[t]{0.25\textwidth}
		\includegraphics[height=0.21\textheight, width=\textwidth]{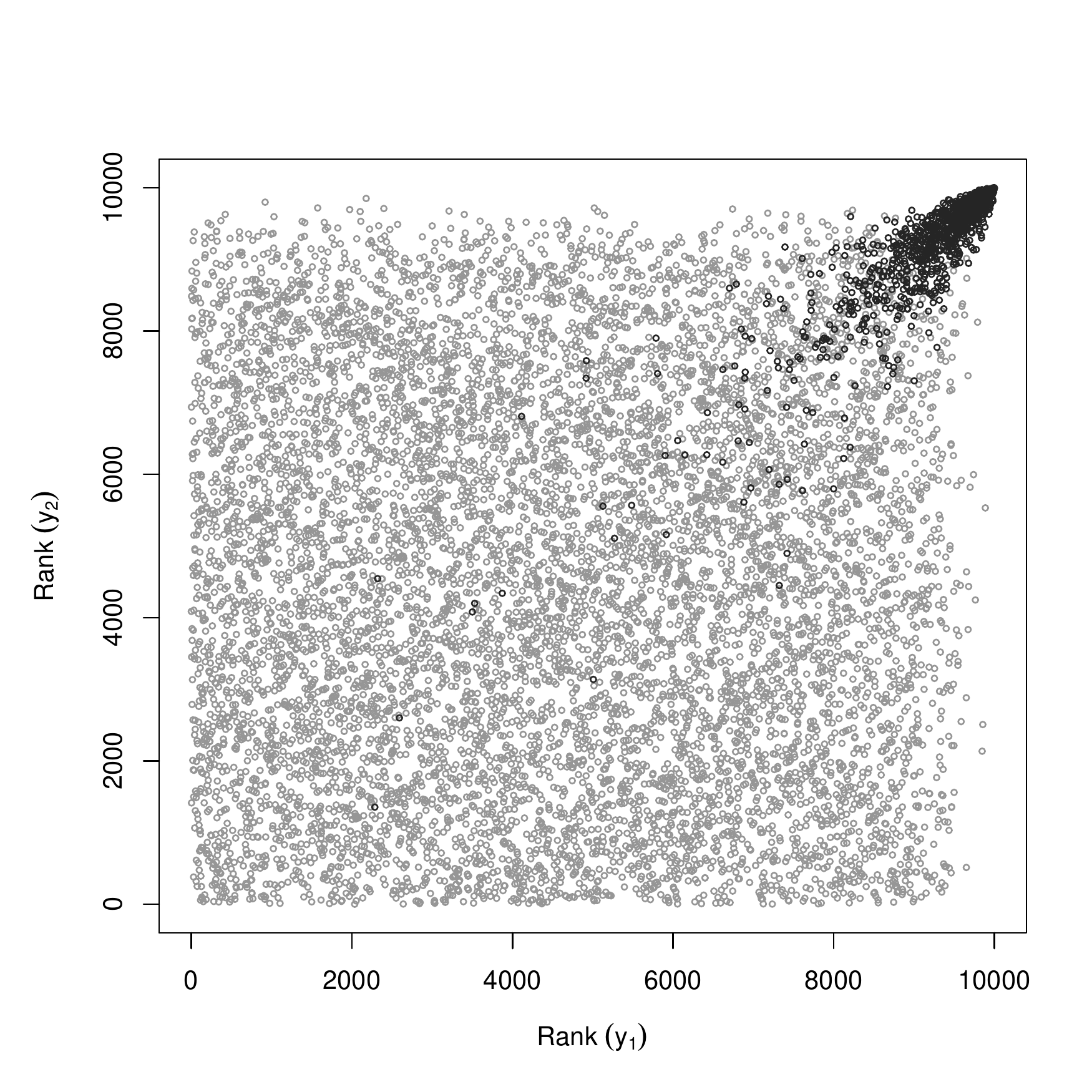}
		\caption{}
	\end{subfigure}
	\hspace{-10pt}
  \begin{subfigure}[t]{0.25\textwidth}
		\includegraphics[height=0.21\textheight, width=\textwidth]{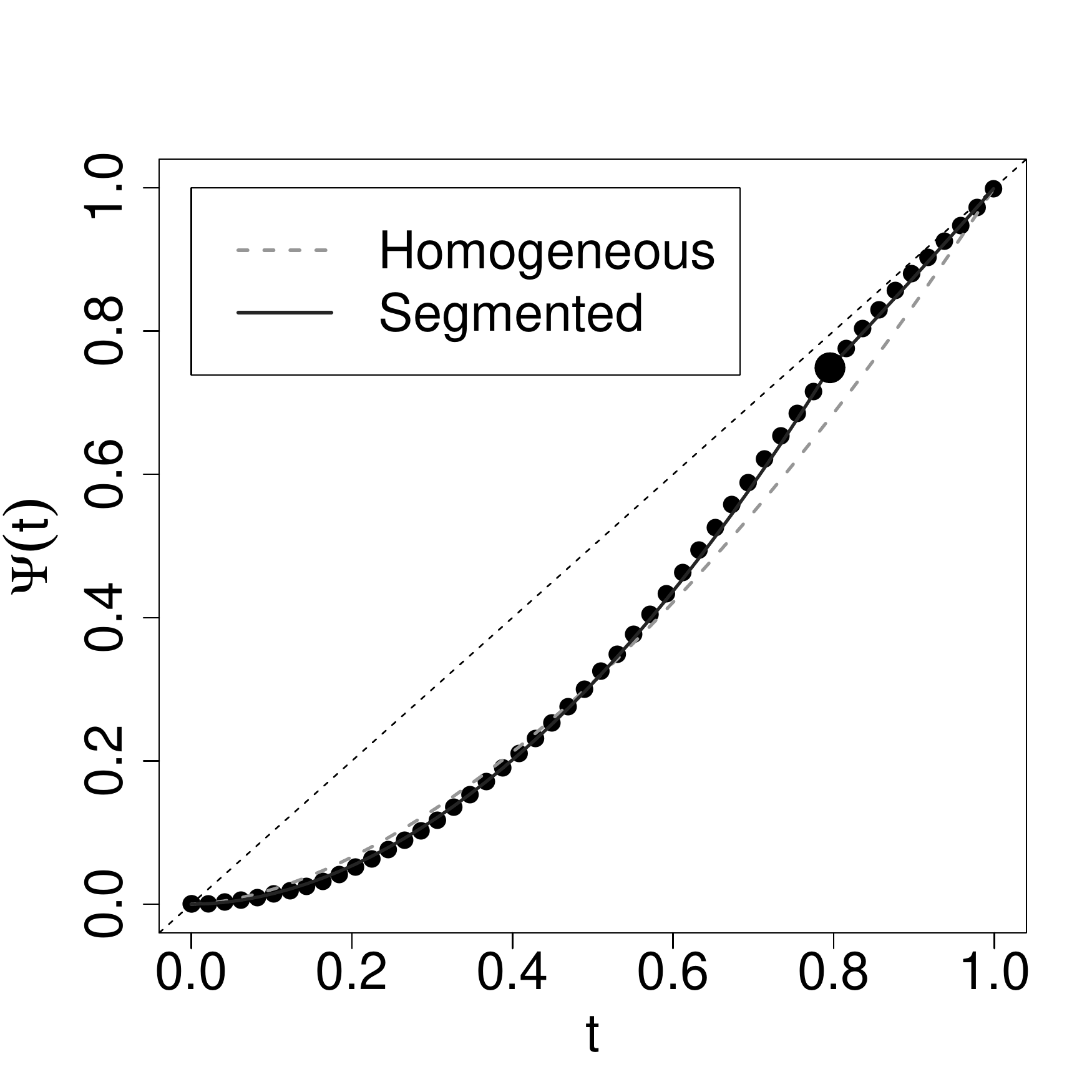}
		\caption{}
	\end{subfigure}
   \hspace{-10pt}
	\begin{subfigure}[t]{0.25\textwidth}
		\includegraphics[height=0.21\textheight, width=\textwidth]  {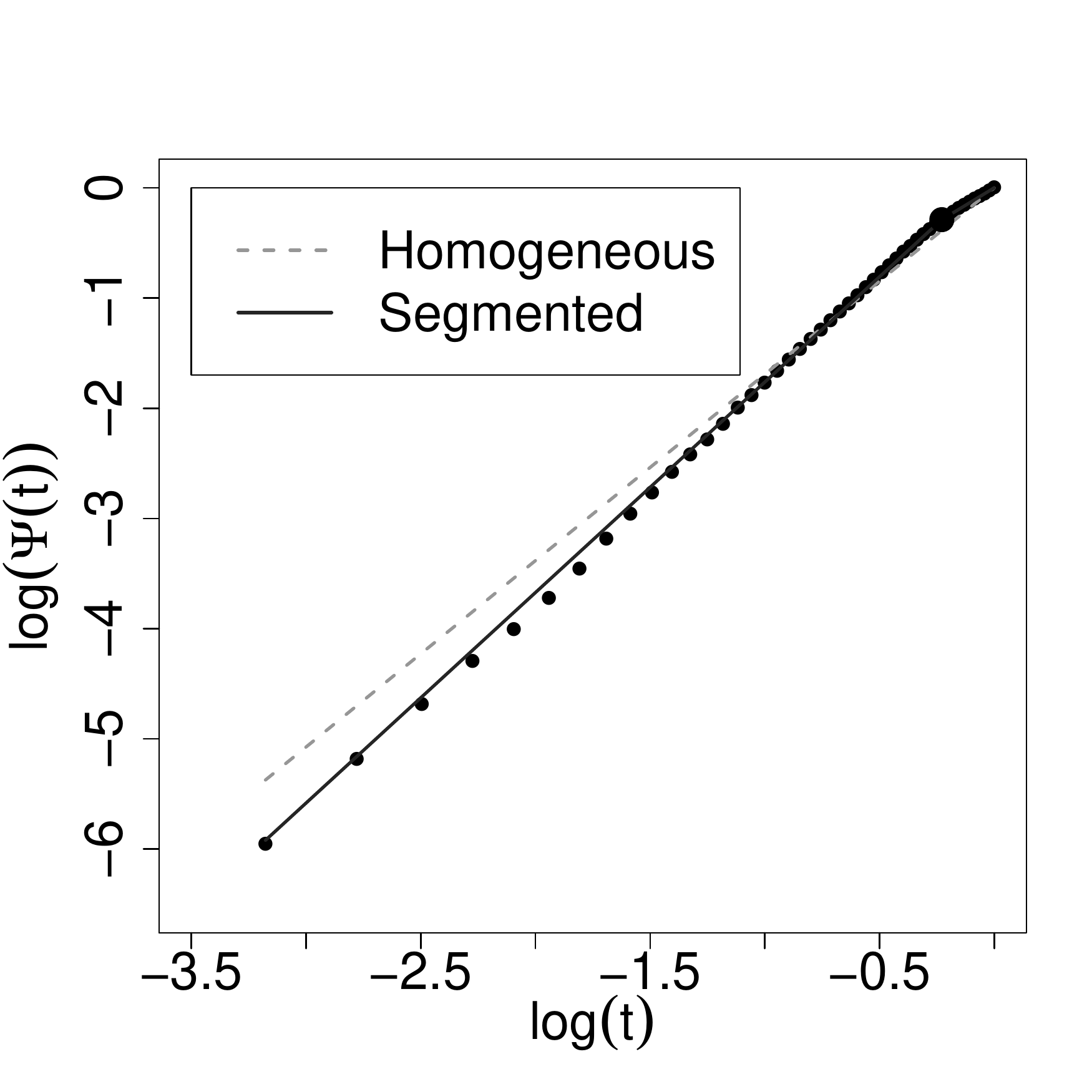}
    	\caption{}
   \end{subfigure}
   \hspace{-10pt}
	\begin{subfigure}[t]{0.25\textwidth}
		\includegraphics[height=0.21\textheight, width=\textwidth]{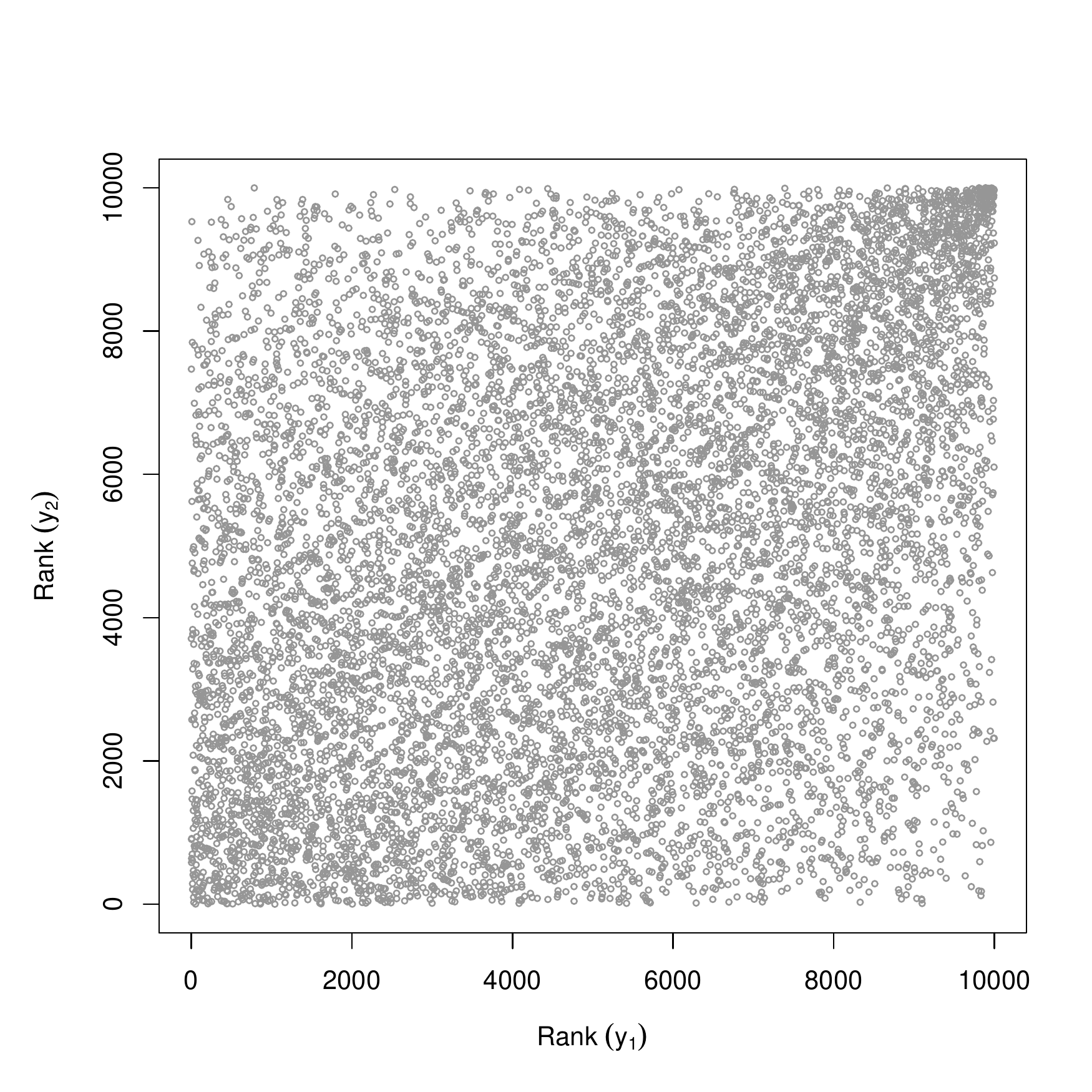}
		\caption{}
	   \end{subfigure}
	\caption{An illustration example for the change of association. 
		(a) Rank scatterplot. Strong candidates (black) receive high ranks and weak ones (gray) receive low ranks. 
		(b) The correspondence curves fitted using homogeneous and segmented models. 
		(c) Exploratory data analysis shows an segmented linear trend between $\log(t)$ and $\log(\Psi(t))$.  
		(d) Data simulated from the fitted homogeneous model \eqref{mod1}. 
		}\label{fig1}
\end{figure}

\begin{figure}
  \centering
     \begin{subfigure}[t]{0.20\textwidth}
	  \includegraphics[height=0.18\textheight, width=1.05\textwidth]  {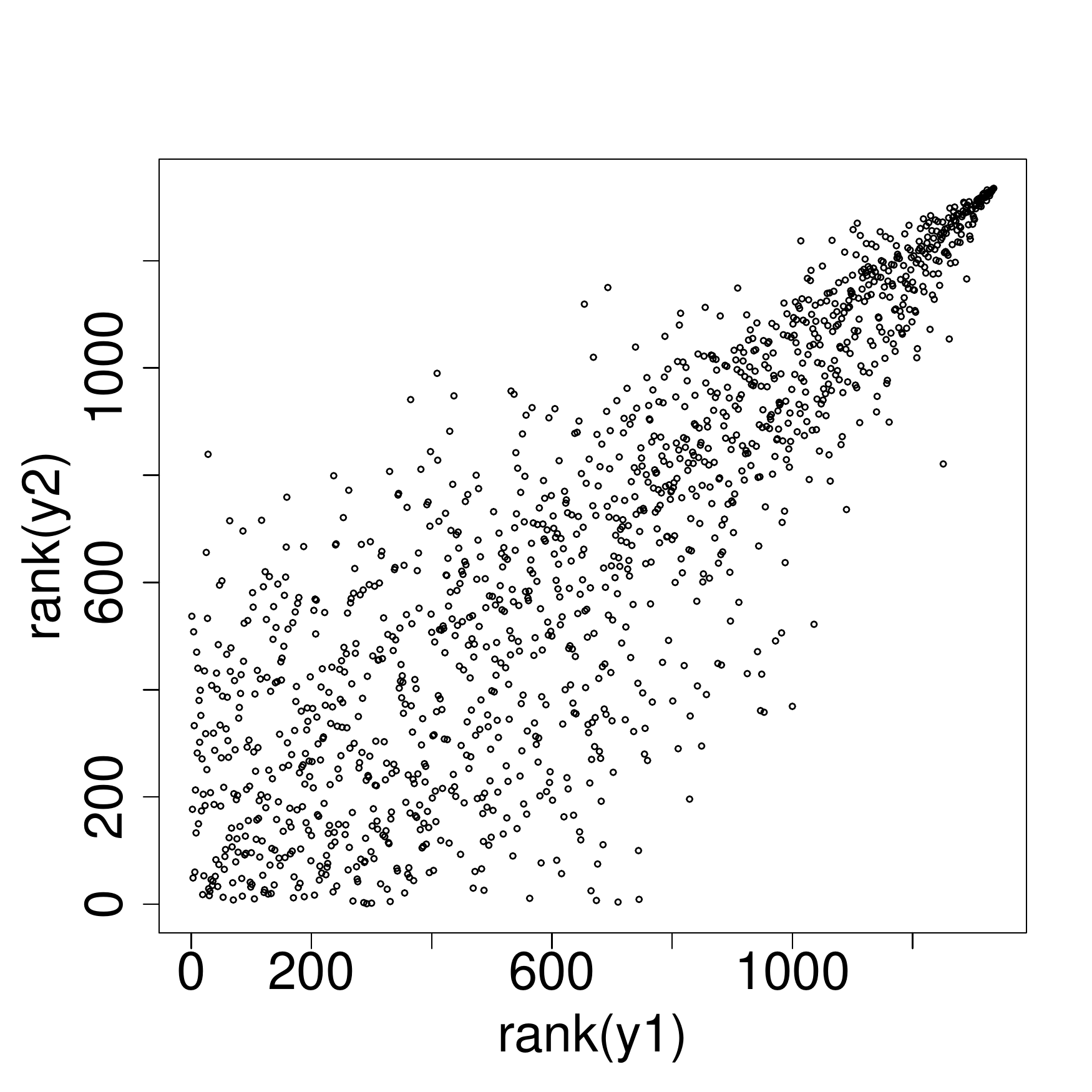}
	    \caption{0.45M}
   \end{subfigure}
   \hspace{-10pt}
   \begin{subfigure}[t]{0.20\textwidth}
	\includegraphics[height=0.18\textheight, width=1.05\textwidth]  {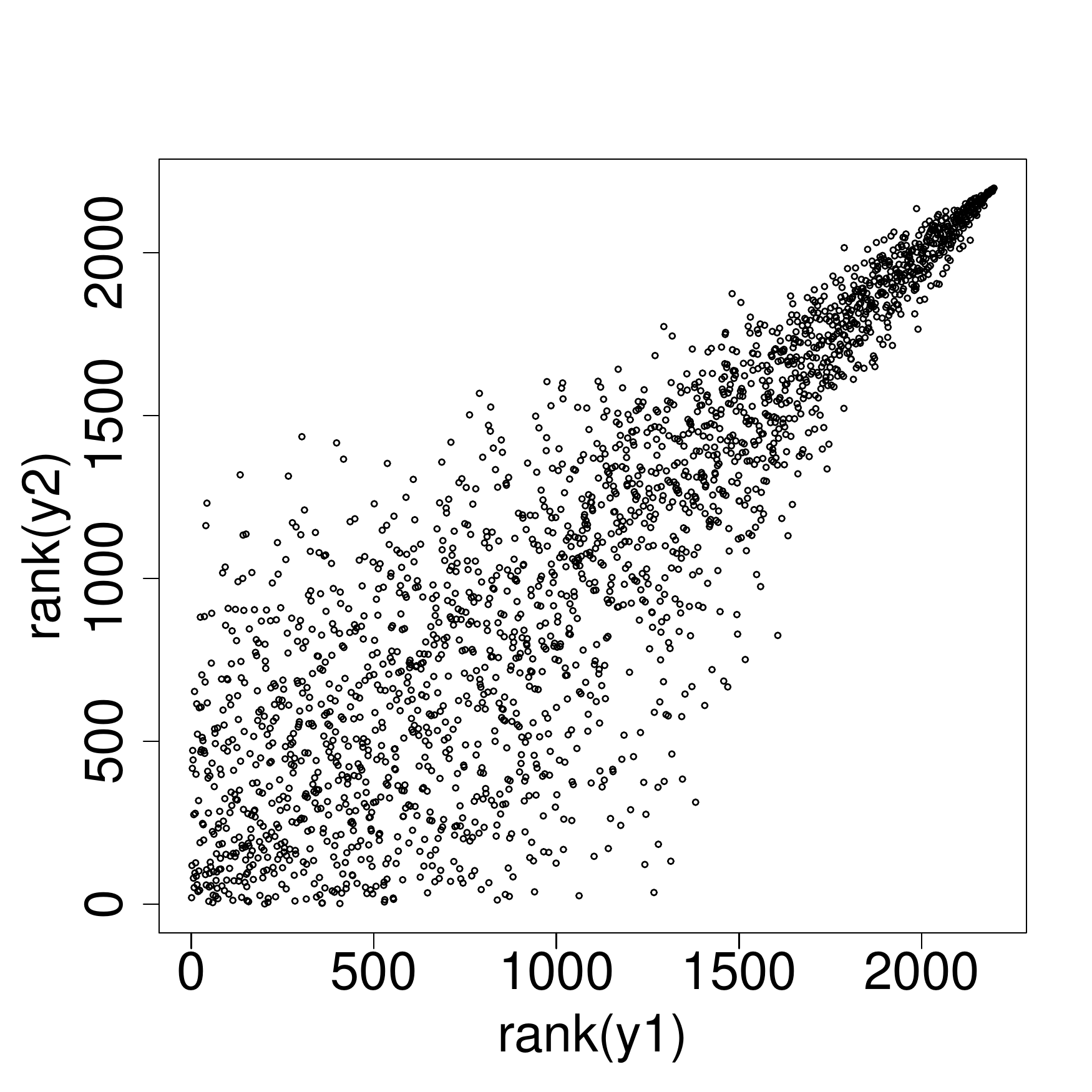}
	\caption{0.9M}
   \end{subfigure}
   \hspace{-10pt}
   \begin{subfigure}[t]{0.20\textwidth}
       \includegraphics[height=0.18\textheight, width=1.05\textwidth]{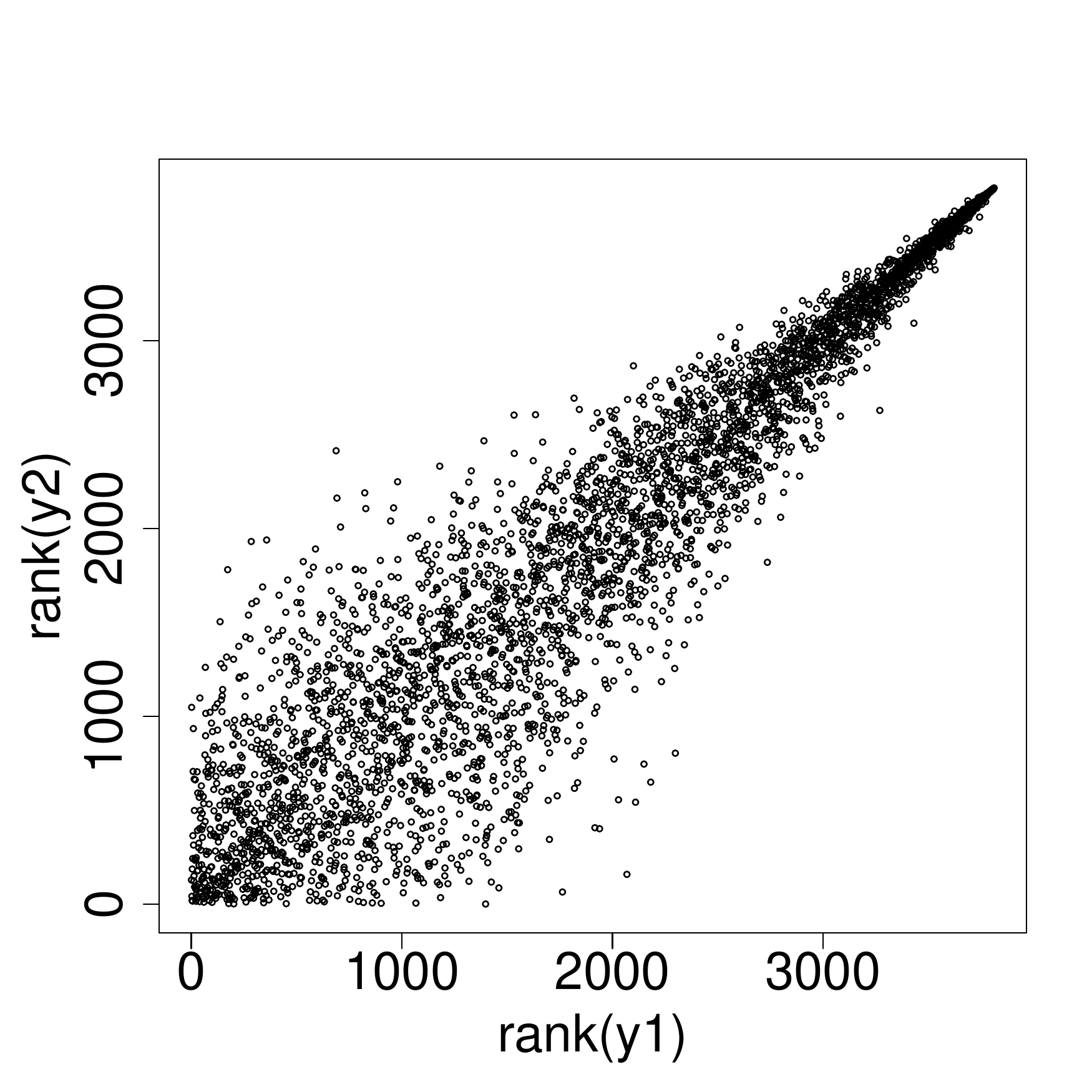}
 	    \caption{2.7M}
   \end{subfigure}
   \hspace{-10pt}
    \begin{subfigure}[t]{0.20\textwidth}
  	 	\includegraphics[height=0.18\textheight, width=1.05\textwidth]{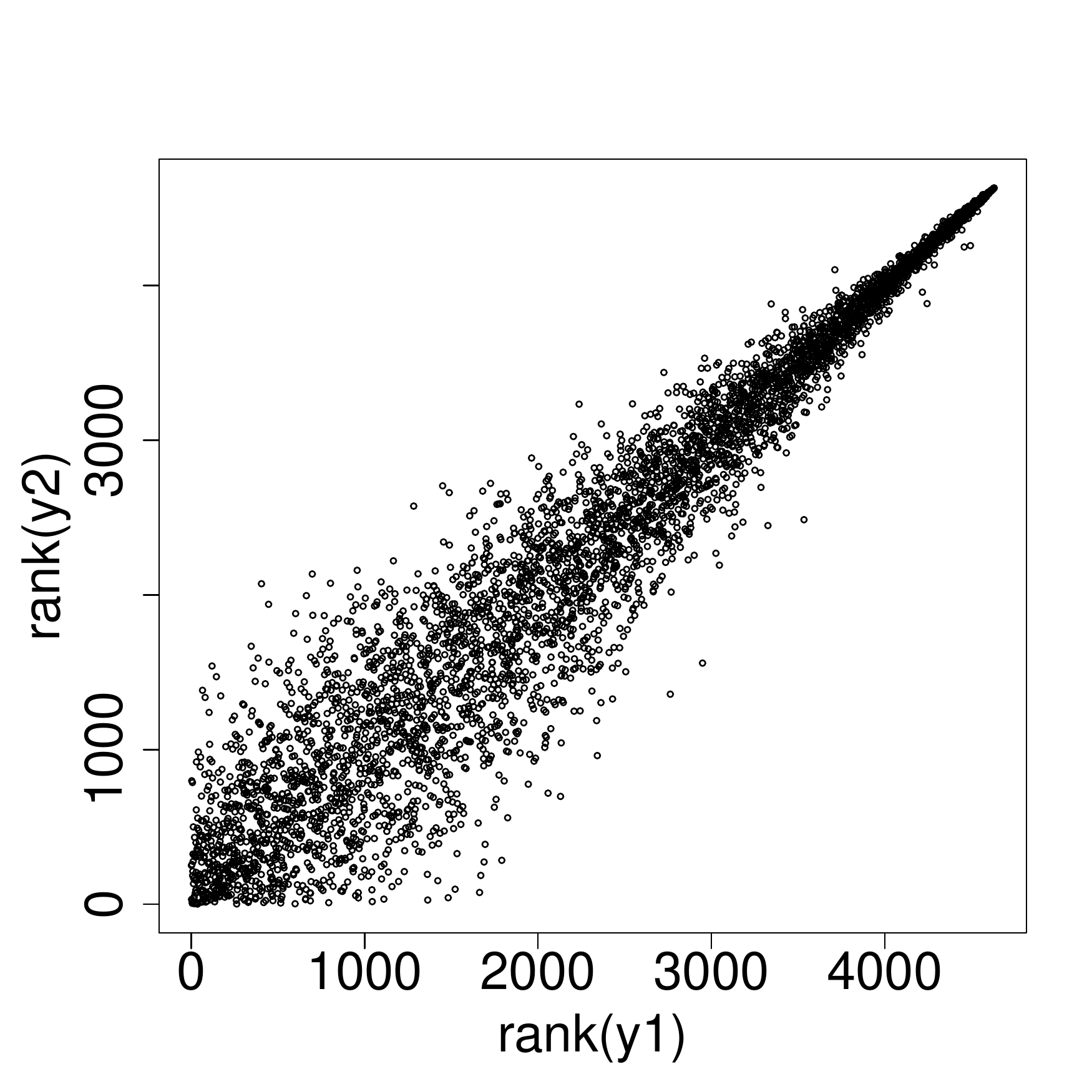}
   		\caption{5.4M}
   \end{subfigure}
   \hspace{-10pt}
	\begin{subfigure}[t]{0.20\textwidth}
		\includegraphics[height=0.18\textheight, width=1.05\textwidth]{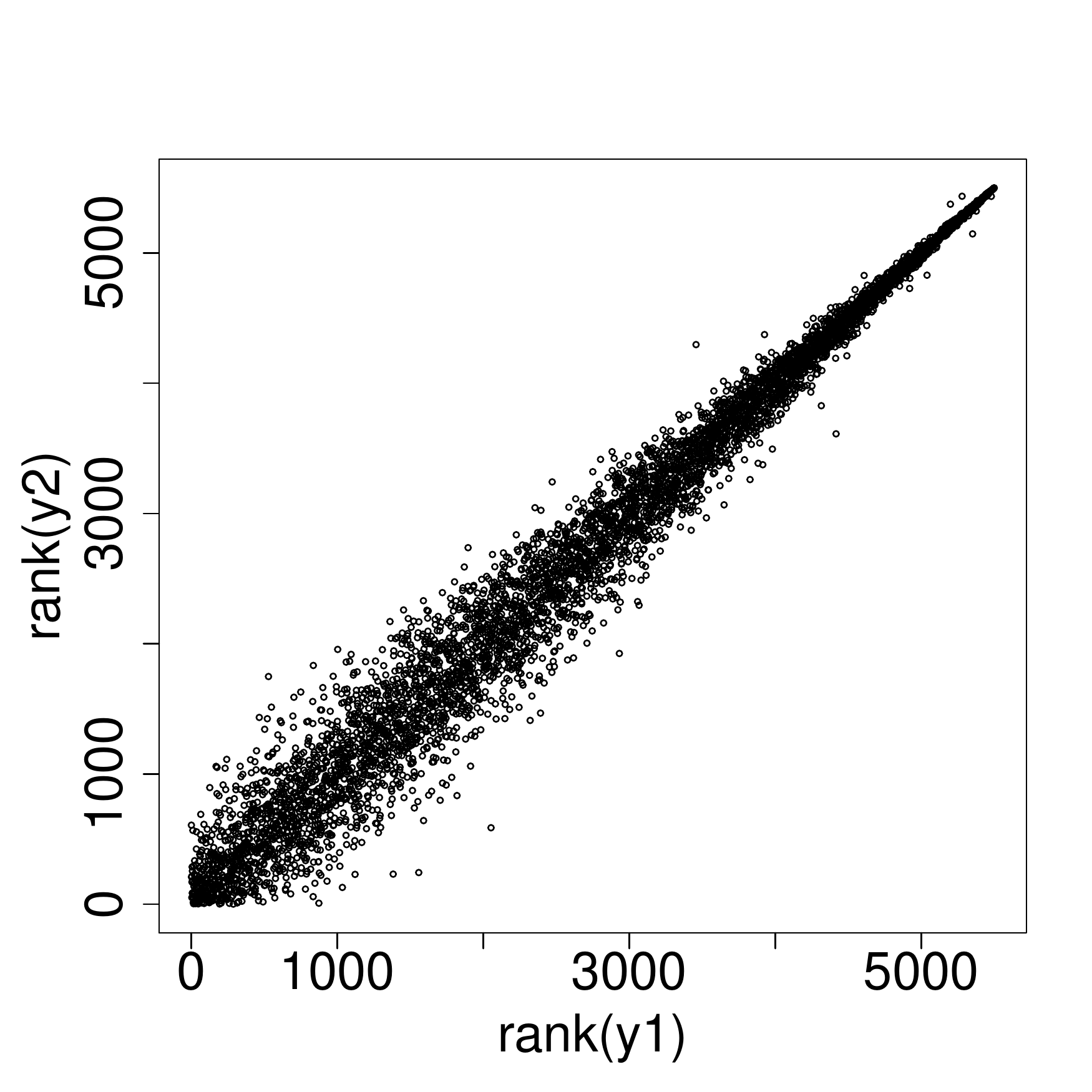}
		\caption{16.2M}
	\end{subfigure}
     \caption{Scores of binding sites in a pair of replicate samples identified from the ChIP-seq data in \citep{chen2012systematic} at different sequencing depths.
     	(a)--(e) Rank scatterplots at depths 0.45M, 0.9M, 2.7M, 5.4M, 16.2M, respectively.
     }
     \label{fig:ChIP}
\end{figure}

Note that the correspondence curve regression model (\ref{E:glm}) has a constant $\alpha_k$ in the entire range of $t$. 
Essentially, this assumes that the change of reproducibility ($g(\Psi(t))$) with the threshold ($h(t)$) is constant across all the thresholds, i.e. the so-called homogeneous reproducibility assumption in \citep{li2017regression}.   
However, this assumption is likely to be violated when strong and weak candidates have considerablely different reproducibility.
As an illustration, 
Figure~\ref{fig1}a  shows  a pair of candidate lists with 20\% strong candidates and 80\% weak ones, simulated from a 2-component Gaussian mixture model.  
A distinct inflection point can be seen on the correspondence curve   (Figure~\ref{fig1}b) at the transition  between noise and signal ($t=0.8$).
An attempt to fit the homogeneous model (\ref{mod1}) shows that there is a clear segmented relationship  between $g(\Psi(t))$ and  $h(t)$ (Figure~ \ref{fig1}c), due to the violation of the homogeneous reproducibility assumption. 
A considerable amount of departure is also observed between the original data and the data simulated from the fitted homogeneous model (Figure~\ref{fig1}d) for the strong candidates.

This motivates us to develop a segmented model for correspondence curve regression, in order to obtain a more accurate description of the covariate effects on strong and weak candidates and simultaneously estimate the onset of the transition.  
Specifically, we model the correspondence curve as two segments that intersect at a change-point $\tau$:  
\begin{align}\label{mod2}
g\left(\Psi(t)\right) =\displaystyle
\begin{cases}
a_1 +b_1 h(t), & t\leq \tau \\
a_2 +b_2 h(t), & t> \tau,
\end{cases}
\end{align} 
where $\tau$ is an unknown change-point and $g(\cdot)=h(\cdot)$ are known functions, which can be determined as in  \citep{li2017regression}.
As $\Psi(t)$ is continuous at $\tau$, 
$a_1 + b_1 h(\tau) = a_2 + b_2 h(\tau)$. 
Then $a_2 =a_1-(b_2-b_1)h(\tau)$. 
Thus, the model \eqref{mod2} can be re-expressed as 
\begin{align}\label{mod3}
\begin{aligned}
g\left(\Psi(t)\right) 
&=\displaystyle
\begin{cases}
a_1 +b_1 h(t), &	t\leq \tau \\
a_1 -(b_2-b_1)h(\tau)+ b_2 h(t), & t> \tau
\end{cases}\\
& = a_1 + b_1 h(\tau) + 
b_1 \left[h(t)-h(\tau)\right]_{-} + 
b_2 \left[h(t)-h(\tau)\right]_{+}\\
& \equiv
\beta_0 + \beta_1 h(\tau) + 
\beta_1 \left[h(t)-h(\tau)\right]_{-} + 
\beta_2 \left[h(t)-h(\tau)\right]_{+}, 
\end{aligned}
\end{align} 
where $a_- = \min(a,0)$ and $a_+=\max(a, 0)$. According to the definition of $\Psi(\cdot)$, $\Psi(\cdot)$ should satisfy $\Psi(0)=0$ and $\Psi(1)=1$, as in the correspondence curve regression model.  

Throughout this work, we will focus on the regression form for the Gumble-Hougaard copula (\ref{mod3}), as this copula provides a reasonable fit for the ranked candidate lists from many high-throughput experiments and its corresponding regression form is simple. 
After plugging in its functional form (i.e. $g(\cdot)=h(\cdot)=\log(\cdot)$) and incorporating the constraints on $\Psi(\cdot)$,
the model \eqref{mod3} reduces to an unconstrained model 
\begin{equation} \label{mod_aim}
\log\left(\Psi(t)\right) = \beta_1 \left[\log(t)-\log(\tau)\right]_{-} + 
\beta_2 \left\{ \log(\tau) + \left[\log(t)-\log(\tau)\right]_{+} \right\}.
\end{equation}
This simplifies the estimation procedure.  Thus we will use (\ref{mod_aim}) hereafter. 
To assess the influence of operational factors on the reproducibility of workflows, we incorporate the operational factors in the model (\ref{mod_aim}) as covariates  
$\bx = (x_0, x_1, \ldots, x_S)^\top$,  
\begin{align}\label{mod4}
\begin{aligned}
\log\left(\Psi(t)|\bx\right) 
& = \sum_{s=0}^S \left[\beta_{s1} \left[\log(t)-\log(\tau)\right]_-  
+ \beta_{s2}\left\{\log(\tau)+\left[\log(t)-\log(\tau)\right]_+\right\}\right]x_s, \\
&\equiv
\sum_{s=0}^S  x_s\bbeta_s^\top W(t, \tau), 
\end{aligned}
\end{align} 
where  $x_0\equiv 1$ for baseline, 
$\bbeta_s=(\beta_{s1}, \beta_{s2})^\top (s=0, \ldots, S)$ are unknown regression coefficients, $\tau$ is an unknown change point, and $W(t, \tau)=\left(\left[\log(t)-\log(\tau)\right]_-, \log(\tau)+\left[\log(t)-\log(\tau)\right]_+ \right)^\top$. If a change of effects on reproducibility is present, then $\beta_{s1} \neq \beta_{s2} ~(s=0,\cdots,S)$. 
Because the Gumble-Houggard copula has a higher association in its upper tail than its lower tail \citep{nelsen2006introduction}, it requires the higher ranked candidates be the more reproducible group. 
We order the rank lists in this way, such that strong candidates consititute the upper segment ($t>\tau$) and weak candidates the lower segment ($t\leq \tau$). 
Accordingly, $\beta_{s1}$ represents the covariate effects for weak candidates and $\beta_{s2}$ represents that for strong candidates.

\subsection{Estimation of regression parameters}   
We use maximum likelihood estimation to estimate the parameters ($\bbeta=(\bbeta_0^\top, \ldots, \bbeta_S^\top)^\top$ and $\tau$) in the model \eqref{mod4}.
To compute the likelihood function  for the model \eqref{mod4}, 
we partition the sample space into non-overlapping categories using the cutoffs and then compute the likelihood based on the multinomial distribution of the categories, 
as in \citet{li2017regression}. 
For fixed $M$, 
let $0=t_0<t_1<\cdots<t_M=1$ be the cutoffs. 
For scores $(Y_{i1}^s, Y_{i2}^s)_{i=1}^{n}$ from the workflow $s$,  
the categories can be defined as 
$$
\mathbb{Y}_m^s=\left\{i:\mathop{\sum}\limits_{j=1}^2 I_{ij}^s(t_m)=2\right\}\backslash
\left\{i:\mathop{\sum}\limits_{j=1}^2 I_{ij}^s(t_{m-1})=2\right\}, 
$$
where $I_{ij}^s(t)=I(Y_{ij}^s\leq \mathbb{F}_j^{s^{-1}}(t))$. 
Let $U_{im}^s= I(i\in \mathbb{Y}_m^s|\bx^s)$
be the binary indicator that the response for the candidate $i$ is in the $m$th category. Denote $\btheta=(\bbeta_0^\top,\ldots, \bbeta_S^\top,  \tau)^\top$,
then the likelihood function for model \eqref{mod4} is
\begin{align}\label{lik1}
\begin{aligned}
\ell_n(\bm{\theta})
&=
\ell_n(\bbeta, \tau)
= \sum_{i=1}^{n}\sum_{s=0}^S\sum_{m=1}^M 
U_{im}^s \log \left\{P(i\in \mathbb{Y}_m^s|\bx^s)\right\}\\
&=
\sum_{i=1}^{n}\sum_{s=0}^S\sum_{m=1}^M
U_{im}^s \log \left\{
\exp\left(\sum_{s=0}^S  x_s\bbeta_s^\top W(t_{m}, \tau)\right)
-\exp\left(\sum_{s=0}^S  x_s\bbeta_s^\top W(t_{m-1}, \tau)\right)
\right\},  
\end{aligned}
\end{align}

Because of the involvement of the change point $\tau$, the log-likelihood \eqref{lik1} is not differentiable at $\tau$, making 
it difficult to maximize directly using common gradient optimization algorithms. 
To overcome this issue, we adopt a grid search strategy. 
This approach is commonly used to solve change-point problems,  
similar to the idea of the profile likelihood. 
Specifically,  
for a fixed $\tau \in (0, 1)$, the parameters $\bbeta$ can be estimated by maximizing the log-likelihood function 
\[
\widehat{\bbeta}(\tau) =\arg\max_{\bbeta} \ell_n(\bbeta, \tau). 
\]
Then, $\tau$ can be obtained by maximizing  $\ell_n\left(\widehat{\bbeta}(\tau), \tau\right)$ over $\tau\in (0, 1)$,  i.e.  
\[
\widehat{\tau} =\arg\max_\tau \ell_n \left(\widehat{\bbeta}(\tau), \tau\right).
\]
The estimate for $\btheta$ is given by 
$\widehat{\btheta} = 
\left(\widehat{\bbeta}(\widehat{\tau}), \widehat{\tau}\right)$.

\subsection{Asymptotic properties }  

We derive the asymptotic properties of $\widehat{\btheta}$ using the modern empirical processes theory.  
For notational simplicity, 
let  
 $\Delta \left(f(t, \mathcal{Z})\right) \equiv f(t_m, \mathcal{Z})- f(t_{m-1}, \mathcal{Z})$ denote the difference of a function $f$ and $\otimes$ the Kronecker product of two matrices.   
Using the notation of empirical processes, we write the log-likelihood \eqref{lik1} and its expectation as 
\[
\ell_n(\btheta) =\mathbb{P}_n m_{\btheta} 
\quad \text{and} \quad 
\ell(\btheta) = \mbox{P}m_{\btheta}, 
\]
respectively, 
where $\mathbb{P}_n=n^{-1}\mathop{\sum}\limits_{i=1}^n \delta_{\mathcal{Z}_i}$ is the empirical measure for  the observed data $\mathcal{Z}_i \equiv (\bY_i, \bx_i)$,  
\begin{align*}
\begin{aligned}
m_{\btheta}(\mathcal{Z})
&=
\sum_{m=1}^M \sum_{s=0}^S
U_{im}^s
\log\left[ 
\exp\left\{\sum_{s=0}^S x_s\bbeta_s^\top W(t_m, \tau) \right\}
-
\exp\left\{\sum_{s=0}^S x_s\bbeta_s^\top W(t_{m-1}, \tau) \right\}
\right], \\
&\equiv 
\sum_{m=1}^M \sum_{s=0}^S
U_{im}^s
\log\left[ 
\Delta \left(
\exp\left\{\sum_{s=0}^S x_s\bbeta_s^\top W(t_m, \tau) \right\}
\right)
\right]. 
\end{aligned}
\end{align*}

Define the variance-covariance matrix $\Sigma(\btheta) \equiv \mbox{E} \dot{m}_{\btheta}\dot{m}_{\btheta}^\top$,  
where $\dot{m}_{\btheta}$ is the derivative of $m_{\btheta}$, 
\[
\dot{m}_{\btheta} = 
\begin{bmatrix}
& \dot{m}_{\bbeta}&\\
& \dot{m}_\tau &
\end{bmatrix}=
\begin{bmatrix}
\mathop{\sum}\limits_{m=1}^M \mathop{\sum}\limits_{s=0}^S
U_{im}^s
\frac{ 
	\Delta \left(\exp\left\{\mathop{\sum}\limits_{s=0}^S x_s\bbeta_s^\top W(t_m, \tau)\right\}\bx_s\otimes W(t_m, \tau) \right)
}
{\Delta \left(\exp\left\{\mathop{\sum}\limits_{s=0}^S x_s\bbeta_s^\top W(t_m, \tau)\right\} \right)
}&\\
\mathop{\sum}\limits_{m=1}^M \mathop{\sum}\limits_{s=0}^S
U_{im}^s
\frac{
	\Delta \left(\exp\left\{\mathop{\sum}\limits_{s=0}^S x_s\bbeta_s^\top W(t_m, \tau) \right\}
	\mathop{\sum}\limits_{s=0}^S(\beta_{s2}-\beta_{s1})\tau^{-1} I(t_m\leq \tau)  \right)
}
{
	\Delta  \left(\exp\left\{\mathop{\sum}\limits_{s=0}^S x_s\bbeta_s^\top W(t_m, \tau) \right\} \right)
}& 
\end{bmatrix}.
\]
The Hessian matrix is defined by 
\[
A(\btheta) \equiv 
\frac{\partial^2}{\partial \btheta \partial \btheta^\top}\ell(\btheta)
=\begin{bmatrix}
\frac{\partial^2}{\partial \bbeta \partial \bbeta^\top}\ell(\btheta) & 
\frac{\partial^2}{\partial \bbeta \partial \tau}\ell(\btheta) \\
\frac{\partial^2}{\partial \tau \partial \bbeta^\top}\ell(\btheta) & 
\frac{\partial^2}{\partial \tau^2}\ell(\btheta)
\end{bmatrix}
=
\begin{bmatrix}
A_{11}(\btheta) & A_{12}(\btheta) \\
A_{21}(\btheta) & A_{22}(\btheta)
\end{bmatrix},
\]
where  
\begin{scriptsize}
	\begin{align*}
	A_{11}(\btheta) 
	&= 
	\mbox{E}\sum_{m=1}^M \sum_{s=0}^{S}
	U_{1m}^s
	\left[
	\frac{
		\Delta \left(\exp\left\{\mathop{\sum}\limits_{s=0}^S x_s\bbeta_s^\top W(t_m, \tau)\right\}
		\bx\otimes W(t_m, \tau)^{\otimes 2}  \right)
	}
	{
		\Delta \left(\exp\left\{\mathop{\sum}\limits_{s=0}^S x_s\bbeta_s^\top W(t_m, \tau)\right\} \right) 
	}
	- \frac{
		\left[\Delta \left(\exp\left\{\mathop{\sum}\limits_{s=0}^S x_s\bbeta_s^\top W(t_m, \tau)\right\} \bx\otimes W(t_m, \tau)\right) \right]^{\otimes 2} 
	}
	{
		\left[\Delta \left(\exp\left\{\mathop{\sum}\limits_{s=0}^S x_s\bbeta_s^\top W(t_m, \tau)\right\}  \right) \right]^2
	}
	\right], \\
	A_{12}(\btheta) 
	&= 
	\mbox{E}\sum_{m=1}^M  \sum_{s=0}^S U_{1m}^s
	\Bigg[
	\frac{
		\Delta
		\left(  
		\exp\left\{\mathop{\sum}\limits_{s=0}^S x_s\bbeta_s^\top W(t_m, \tau)\right\}
		\left[\mathop{\sum}\limits_{s=0}^S\beta_{s2}\tau^{-1}I(t_m\leq \tau) \bx\otimes  W(t_m, \tau)  
		+\bx\otimes \dot{W}(t_m, \tau)\right]
		\right) 
	}
	{
		\Delta 
		\left(  
		\exp\left\{\mathop{\sum}\limits_{s=0}^S x_s\bbeta_s^\top W(t_m, \tau)\right\}
		\right) 
	}\\
	&- \frac{
		\Delta 
		\left( 
		\exp\left\{\mathop{\sum}\limits_{s=0}^S x_s\bbeta_s^\top W(t_m, \tau)\right\}\bx\otimes W(t_m, \tau)
		\right) 
		\cdot
		\Delta 
		\left( 
		\exp\left\{\mathop{\sum}\limits_{s=0}^S x_s\bbeta_s^\top W(t_m, \tau)\right\}
		\mathop{\sum}\limits_{s=0}^S x_s\left(\beta_{s2}-\beta_{s1}\right)\tau^{-1}I(t_m\leq \tau)
		\right) 
	}
	{
		\left[
		\Delta 
		\left( 
		\exp\left\{\mathop{\sum}\limits_{s=0}^S x_s\bbeta_s^\top W(t_m, \tau)\right\}
		\right) 
		\right]^2
	}
	\Bigg], 	\\
	A_{22}(\btheta) 
	&= 
	\mbox{E}\sum_{m=1}^M \sum_{s=0}^S U_{1m}^s 
	\Bigg[
	\frac{
		\Delta
		\left(  
		\exp\left\{\mathop{\sum}\limits_{s=0}^S x_s\bbeta_s^\top W(t_m, \tau)\right\}
		\left[\left(\mathop{\sum}\limits_{s=0}^S x_s(\beta_{s2}-\beta_{s1}) \tau^{-1}I(t_m\leq \tau) \right)^2 
		-\mathop{\sum}\limits_{s=0}^S x_s(\beta_{s2}-\beta_{s1}) \tau^{-2}I(t_m\leq \tau) 
		\right]
		\right)
	}
	{
		\Delta 
		\left(  
		\exp\left\{\mathop{\sum}\limits_{s=0}^S x_s\bbeta_s^\top W(t_m, \tau)\right\}  
		\right) 
	}\\
	&- \frac{
		\left[
		\Delta 
		\left(
		\exp\left\{\mathop{\sum}\limits_{s=0}^S x_s\bbeta_s^\top W(t_m, \tau)\right\}
		\mathop{\sum}\limits_{s=0}^S x_s(\beta_{s2}-\beta_{s1}) \tau^{-1}I(t_m\leq \tau) 
		\right) 
		\right]^2
	}
	{
		\left[
		\Delta 
		\left( 
		\exp\left\{\mathop{\sum}\limits_{s=0}^S x_s\bbeta_s^\top W(t_m, \tau)\right\}
		\right) 
		\right]^2
	}\Bigg].	
	\end{align*}
\end{scriptsize}
and  $A_{21}(\btheta)= A_{12}(\btheta)^\top$, 
where $\dot{W}(t,\tau)=\left(-\tau^{-1}I(t\leq \tau), \tau^{-1}I(t\leq \tau)\right)^\top$ 

\begin{theorem}\label{thm1}
	Under the regularity conditions in the Appendix,   
	$\widehat{\btheta}$ converges to $\btheta$ in probability, as $n$ goes to infinity. 
	Furthermore, 
	$\sqrt{n}(\widehat{\btheta}-\btheta_0)$ 
	is asymptotically normally distributed with  mean zero and covariance matrix 
	$A(\btheta_0)^{-1}\Sigma(\btheta_0) A(\btheta_0)^{-1}$, 
	as $n\rightarrow \infty$. 	
\end{theorem}

The asymptotic variance-covariance matrix can be estimated by $\widehat{A}(\widehat{\btheta})^{-1} \widehat{\Sigma}(\widehat{\btheta}) \widehat{A}(\widehat{\btheta})^{-1}$, 
where $ \widehat{\Sigma}(\widehat{\btheta}) = 
n^{-1}\mathop{\sum}\limits_{i=1}^n \widehat{\dot{m}}_i(\widehat{\btheta}) \widehat{\dot{m}}_i(\widehat{\btheta})^\top$ with 
\[
\widehat{\dot{m}}_i(\widehat{\btheta}) =
\begin{bmatrix}
\mathop{\sum}\limits_{m=1}^M \mathop{\sum}\limits_{s=0}^S
U_{im}
\frac{ 
	\Delta \left(\exp\left\{\mathop{\sum}\limits_{s=0}^S x_s\widehat{\bbeta}_s^\top W(t_m, \widehat{\tau})\right\}\bx\otimes W(t_m, \widehat{\tau}) \right)
}
{\Delta \left(\exp\left\{\mathop{\sum}\limits_{s=0}^S x_s\widehat{\bbeta}_s^\top W(t_m, \widehat{\tau})\right\} \right)
}&\\
\mathop{\sum}\limits_{m=1}^M \mathop{\sum}\limits_{s=0}^S
U_{im}
\frac{
	\Delta \left(\exp\left\{\mathop{\sum}\limits_{s=0}^S x_s\widehat{\bbeta}_s^\top W(t_m, \widehat{\tau}) \right\}
	\mathop{\sum}\limits_{s=0}^S x_s(\widehat{\beta}_{s2}-\widehat{\beta}_{s1})\widehat{\tau}^{-1}I(t_m\leq \widehat{\tau})  \right)
}
{
	\Delta  \left(\exp\left\{\mathop{\sum}\limits_{s=0}^S x_s\widehat{\bbeta}_s^\top W(t_m, \widehat{\tau}) \right\} \right)
}& 
\end{bmatrix},
\]
and $\widehat{A}(\widehat{\btheta})$ is obtained by replacing the population parameters by their empirical counterparts.   
However, the plug-in version of the asymptotic covariance matrix in Theorem~\ref{thm1} 
is computationally complicated.  
In practice, we estimate the standard error of $\widehat{\btheta}$ using a nonparametric bootstrap method.  
Theorem~\ref{thm1}  ensures that the bootstrap method would produce valid estimate of the variance of $\btheta$.

\subsection{Test the existence of a change point.} 

In practice, it is possible that the data does not have a change in reproducibility. For example, if the potential candidates have been screened with a stringent threshold, then most candidates on the candidate lists are relevant to the biological feature and may have similar reproducibility.
If a change point does not exist, then $\tau$ is unidentifiable and its estimation is ill-conditioned. Thus, it is necessary to test the existence of a change point before performing the estimation.  
Here we develop a test for the existence of a change point for the baseline model \eqref{mod_aim},  
$
\log\left(\Psi(t)\right) = \beta_1 \left[\log(t)-\log(\tau)\right]_{-} + 
\beta_2 \left\{ \log(\tau) + \left[\log(t)-\log(\tau)\right]_{+} \right\}.
$
The test can be applied to each workflow individually to ensure identifiability before using the segmented regression.

Many tests have been developed for determining the existence of a change point in regression models, for example, in
mean regression
\citep{andrews1993tests, bai1996testing, hansen1996inference},  
quantile regression 
\citep{qu2008testing, lich2011bent}, 
transformation models 
\citep{kosorok2007inference}, 
time series models 
\citep{chan1993consistency, cho2007testing},  and others. 
However, no analogous tests have been developed in the context of the correspondence curve regression. 
Because our method is a likelihood-based model, we adopt the sup-likelihood ratio test proposed by  \citet{lee2011testing}, which is a general testing framework for threshold effects in regression models, and derive a test for our model based on this framework. 

Specifically, we consider the null and alternative hypotheses    
\[
H_0: \beta_1=\beta_2  \quad \text{for any} \quad \tau\in [0,1]
\quad vs. \quad 
H_1: \beta_1\neq \beta_2  \quad  \text{for some} \quad  \tau \in [0,1].  
\]
Under $H_0$, the model~\eqref{mod_aim} reduces to 
$\log\left(\Psi(t)\right) = \beta_1 \log(t)$, 
which is the correspondence curve regression model proposed by \cite{li2017regression}. We define the maximum of the likelihood function and the corresponding estimator under $H_0$ as
\[
\widetilde{\ell}
\equiv \max_{(\beta_1, \beta_2, \tau): \beta_1=\beta_2} \ell_n(\beta_1, \beta_2,\tau) 
\quad \text{and} \quad 
\widetilde{\beta}_1 \equiv
\mathop{\arg\max}\limits_{(\beta_1, \beta_2, \tau): \beta_1=\beta_2} \ell_n(\beta_1, \beta_2, \tau),
\]
respectively.
Under $H_1$,  the estimators are obtained by searching the grid of $\tau$.
For a given $\tau$,  
let 
$
(\widehat{\beta}_1(\tau), \widehat{\beta}_2(\tau))
=\mathop{\arg\max}\limits_{(\beta_1, \beta_2)}\ell_n(\beta_1, \beta_2, \tau)$. 
We define the maximum of the profile likelihood and the corresponding estimators as
$\widehat{\ell}\equiv \mathop{\max}\limits_\tau \ell_n\left(\widehat{\beta}_1(\tau), \widehat{\beta}_2(\tau), \tau\right)$, 
$\widehat{\tau}=\mathop{\arg\max}\limits_{\tau}\widehat{\ell_n}\left(\widehat{\beta}_1(\tau), \widehat{\beta}_2(\tau), \tau\right)$,
and $(\widehat{\beta}_1, \widehat{\beta}_2)
=\left(\widehat{\beta}_1(\widehat{\tau}), \widehat{\beta}_2(\widehat{\tau})\right)$, respectively. 
The quasi-likelihood ratio statistic is given by  
\[
QLR=n(\widehat{\ell}-\widetilde{\ell}).
\]
Following an analogous argument in \citet{lee2011testing}, 
one can show that the asymptotic distribution of the QLR test is 
\[
\frac{1}{2}\left[\sup_\tau \mathcal{G}^\top(\tau) V(\tau)^{-1}\mathcal{G}(\tau)
- \mathcal{G}_1^\top V_1^{-1} \mathcal{G}_1\right],
\]
where $\mathcal{G}$ is a Gaussian process with the variance function 
\[
V(\tau)=\hbox{E}\left[
\mathop{\sum}\limits_{m=1}^M  U_{1m}
\frac{ 
	\Delta \left(\exp\left\{\bbeta_0^\top W(t_m, \tau) \right\}W(t_m, \tau) \right)
}
{\Delta \left(\exp\left\{ \bbeta_0^\top W(t_m, \tau) \right\} \right)
}
\right]^{\otimes2},
\]
 $\bbeta_0=(\beta_1,\beta_2)^\top$ 
, $W(t, \tau)=\left(\left[\log(t)-\log(\tau)\right]_-, \log(\tau)+\left[\log(t)-\log(\tau)\right]_+ \right)^\top$, 
and 
$\mathcal{G}_1$ and $V_1$ are the first  element of $\mathcal{G}$ and $V(\tau)$, respectively.  

Since the limiting distribution of the test statistic is nonstandard, it is difficult to obtain the critical values directly. Thus we use a simulation-based procedure to obtain the sampling distribution of the test statistic, as in \citet{lee2011testing}.
The procedure is summarized in Algorithm 1 (Appendix B in Supplementary material).

\section{Simulation Studies} \label{sec:numerical}

In this section, we assess the performance of the proposed method using  simulations resembling the output from high-throughput experiments. In particular,
we evaluate the accuracy of estimation, the quality of model fitting,  and the power for detecting differences in reproducibility between different workflows. 

\subsection{Accuracy of estimation and model fitting for the baseline model}\label{SS:baseline}
To evaluate the accuracy of the estimation procedure, we first consider the baseline model.  To simulate from the regression model, we take advantage of the connection between the correspondence curve regression model (CCR) and Archimedean copulas  in \citet{li2017regression} (see Section \ref{SS:CCR}). 
 That is, for the CCR models with the copula connection, the data simulated from its corresponding Archimedean copula follows the relationship described in the CCR model.  
 Note that, although the segmented model (\ref{mod4}) itself is not a correspondence curve regression model, each of its segments is. 
 Therefore, one may approximate the segmented model using a 2-component mixture model with each component representing one segment. 
 For each component, the correspondence curve regression can be simulated using its corresponding Archimedean copula. The two components then can be joined using a marginal mixture model to form the segmented model.  
 The mixing proportion is equal to the location of the change point.  
 
Note that, this simulation is only an approximation of the segmented model. 
Unlike the segmented model, where the supports of the two segments do not overlap, the two components in the mixture model have overlapping supports. 
Therefore, the effects on one component may spill over into the other component due to the overlap. However, this spillover is quite common in reality, as signal and noise in real data always have some overlap.
In the simulation, the level of spillover can be controlled by the distance between the marginal means of the two components. 

Here we consider two scenarios.  Scenario I corresponds to the situation that the simulated data closely follows the segmented regression model. Scenario II imitates the output from real ChIP-seq experiments and is not generated from the regression model. Because it is not based on the regression model, this analysis also serves as an evaluation of the robustness of our model against model misspecification.

\begin{itemize}
	\item 
	\textbf{Scenario~I}
	In this simulation, we generate each segment from a Gumbel-Hougaard copula  and join them through a marginal Gaussian mixture model, 
\[
	(Y_{i,1}, Y_{i,2}) \sim
	\pi_1 G_{\theta_{1}}\left(
	N(\mu_1,1), N(\mu_1,1)
	\right)
	+
	(1-\pi_1) G_{\theta_{2}}\left(
	N(\mu_2,1), N(\mu_2,1)
	\right),
\]	
where the first component represents the weak candidates and the second one represents the strong ones, $G_{\theta}(\cdot, \cdot)$ is a Gumbel-Hougaard copula with parameter $\theta$, and $N(\mu,\sigma^2)$ is a normal distribution with mean $\mu$ and variance $\sigma^2$.  		
	This simulation approximates the segmented regression model (\ref{mod_aim}).
	We assume that the scores of weak candidates are independent across replicates 
	($\theta_1=1$) and that the scores of strong ones are 
	positively associated across replicates ($\theta_2=1.2$ or $2$).  
As the strong component tends to have higher ranks than the weak one, we set $\mu_1=0$ and $\mu_2=3$. 
The performance is evaluated at the mixture proportions, $\pi_1=0.6,~0.8,~0.9,$ and $~0.95$.   

	\item 
	\textbf{Scenario~II}.  In this simulation, we mimic the ranked lists of identified protein binding sites from a pair of replicate ChIP-seq experiments. As shown in \cite{li2011measuring}, the dependence structure of the ranked lists can be reasonably approximated by a bivariate normal mixture distribution. Therefore, we generate the scores from a bivariate normal mixture distribution,
	$$
	(Y_{i,1}, Y_{i,2}) \sim
	\pi_1 N\left(
	\begin{bmatrix}
	\mu_1\\
	\mu_1\\
	\end{bmatrix},
	\begin{bmatrix}
	1 & \theta_1\\
	\theta_1 & 1\\
	\end{bmatrix}\right)
	+
	(1-\pi_1)
	N\left(
	\begin{bmatrix}
	\mu_2\\
	\mu_2\\
	\end{bmatrix},
	\begin{bmatrix}
	1 & \theta_2\\
	\theta_2 & 1\\
	\end{bmatrix}\right),
	$$
	where the first component represents the weak candidates and the second one represents the strong candidates. 
For weak candidates, we consider two cases, one in which the scores of weak candidates are uncorrelated ($\theta_1=0$) and one in which they are correlated ($\theta_1=0.4$). The latter can arise when the workflow has a systematic bias that makes the weak candidates correlated. As the strong candidates tend to be more consistent and higher ranked than the weak ones, we set $\theta_2= 0.9$, $\mu_1=0$ and $\mu_2=2.5$. We choose a smaller $\mu_2$ than that in Scenario I to allow more overlap between components. 
The performance is evaluated at $\pi_1=0.6,~0.8,~0.9$, and $~0.95$. 
\end{itemize}


For each parameter setting,  we simulate 100 datasets,  each of which consists of $n=10,000$ pairs of observations. For both scenarios, we fit the model (\ref{mod_aim}).   As a comparison, we also fit the data using the estimator in \citet{li2017regression} (denoted as ``Homogeneous'').  We fit each model with $M=100$ equally spaced cutoffs in $(0,1)$ and compare how well each model fits the data. 
To assess model fitting, we evaluate the deviation between the empirical correspondence curve and the fitted curve by computing the mean integrated squared errors for $\Psi(t)$, 
\[
\text{MISE}\left(\widehat{\Psi}(t)\right)
=\int_0^1 \left[\widehat{\Psi}(t)-\widehat{\Psi}_n(t)\right]^2dt, 
\]
where $\widehat{\Psi}(t)$ and $\widehat{\Psi}_n(t)$ are the fitted and empirical correspondence curve, respectively.

As shown in Table~\ref{tab:mise}, 
	the estimated change-point $\tau$ accurately approximates the mixing proportion $\pi_1$ in all simulations.  
The segmented model consistently shows a smaller MISE than the homogeneous model.  The decrease in MISE is especially obvious when the data is not dominated by one component ($\pi_1=0.6$ or $0.8$). 
This is because the data in these cases are more heterogeneous than those with a dominating group ($\pi_1=0.9$ or $0.95$), making the homogeneous model fit poorly. Despite the model misspecification in Scenario~II, our model performs well in both parameter estimation and model fitting, indicating that our model is reasonably robust to model misspecification.  Figure~\ref{fig1} displays the fitted correspondence curve for Scenario~II with $\pi_1=0.8$ and $\theta_1=0$.

\begin{table}
	\centering
	\caption{Estimation of change point and comparison of model fitting in the simulation of the baseline model. Segmented: the proposed model; Homogeneous: the model in \citep{li2017regression}. Model fitting is evaluated by mean integrated squared errors (MISE$\times 10^{-4}$). }
\label{tab:mise}
\scalebox{0.9}{
	\begin{tabular}{c cccccc c cccccc}
	\hline\hline
	\multicolumn{1}{c}{}&
	\multicolumn{6}{c}{Scenario I} && \multicolumn{6}{c}{Scenario II}\\
                  \cline{2-7}   \cline{9-14}\noalign{} 
	\multicolumn{1}{c}{} & 
	\multicolumn{2}{c}{} & \multicolumn{2}{c}{Segmented} &&  \multicolumn{1}{c}{Homogeneous} &&
	\multicolumn{2}{c}{} & \multicolumn{2}{c}{Segmented} &&  \multicolumn{1}{c}{Homogeneous}
	\\ 
                  \cline{4-5} \cline{7-7}  \cline{11-12} \cline{14-14}\noalign{}  
 \multicolumn{1}{c}{$\pi_1$}&
    \multicolumn{1}{c}{$\theta_1$} &
  \multicolumn{1}{c}{$\theta_2$} &	
 \multicolumn{1}{c}{$\widehat{\tau}$ } & 
 \multicolumn{1}{c}{MISE} && 
 \multicolumn{1}{c}{MISE} &
 &  \multicolumn{1}{c}{$\theta_1$} &
  \multicolumn{1}{c}{$\theta_2$} & 
 \multicolumn{1}{c}{$\widehat{\tau}$ } & 
 \multicolumn{1}{c}{MISE} && 
 \multicolumn{1}{c}{MISE} \\
		\hline
0.60& 1.0&	 1.2 & 0.557 & 8.592 &&	17.288  & &
 0.0&	0.9 & 0.596 & 5.320   &&   20.567    \\
0.80&	&	 & 0.773 & 5.487 && 14.969    &&
   &	 & 0.800  & 5.313   &&  15.377     \\
0.90 &		&	 &0.882  & 5.142 &&  9.201    &&
 &	 &  0.903&   5.223   &&   8.960    \\
0.95	&	&   & 0.938&	  5.068 && 6.363 &&
	& 	& 0.955  &   5.125    &&   6.200\\
 \hline
 0.60 & 1.0&	3.0 & 0.584 & 5.280 && 23.072 &&
  0.4&	0.9 & 0.601 & 5.133   &&  10.758  \\
 0.80 &   &	 & 0.796 & 5.236 && 17.750  &&
   & & 0.817  & 5.079   &&  8.118 \\
 0.90&	& & 0.898 & 5.152 && 9.953  &&
 	&	 &0.919  & 5.156      &&  5.772   \\
0.95	&	& &0.953 & 5.115 && 6.547  &&
	&	&  0.966 &  5.280    &&  5.342  \\
   \hline
	\end{tabular}
	}
\end{table}

\subsection{Type I error and power for detecting difference in reproducibility across workflows}

Next, we assess the estimation of the covariate effects on the reproducibility of high-throughput experiments.  
We consider a setting with two workflows, $s=0, 1$, where the workflow 1 ranks strong candidates more reproducibly across replicates than the workflow 0, and both workflows rank weak candidates with the similar level of reproducibility. 
 For each workflow, we consider the same two scenerios as in Section \ref{SS:baseline}. We evaluate the type I error and the power of our method for detecting the difference in reproducibility.

\begin{itemize}
	\item 
	\textbf{Scenario~I with covariates}\\
	Similar to Scenario~I in Section \ref{SS:baseline}, we generate the scores of a pair of replicates from a workflow from
	$$
	\left(Y^s_{i,1}, Y^s_{i,2}\right) \sim
	\pi_1 G_{\theta_{s1}}\left(
	N(\mu_1,1), N(\mu_1,1)
	\right)
	+
	(1-\pi_1) G_{\theta_{s2}}\left(
	N(\mu_2,1), N(\mu_2,1)
	\right),   \quad \text{for $s=0,1$}.
	$$
	To evaluate type I error and power, we consider two cases, one with identical reproducibility in the two workflows ($\theta_{02}=\theta_{12}=1.2$) and one in which the reproducbility in the workflow 1 is higher than that in the workflow 0 for relevant candidates ($\theta_{02}=1.2$ and $\theta_{12}=3$).
	In both cases, we set all the other parameters identical between the workflows, with $\mu_1=0$, $\mu_2=3$,  $\theta_{01}=\theta_{11}=1$, and $\pi_1=0.6$, $0.8$, $0.9$, and $0.95$, as in Section \ref{SS:baseline}.   
		
	\item 
	\textbf{Scenario~II with covariates}\\
	Similar to Scenerio~II in Section \ref{SS:baseline}, we generate the scores of a pair of replicates from each workflow from
	\[
	\left(Y^s_{i,1}, Y^s_{i,2}\right) \sim
	\pi_1 N\left(
	\begin{bmatrix}
	\mu_1\\
	\mu_1\\
	\end{bmatrix},
	\begin{bmatrix}
	1 & \theta_{s1}\\
	\theta_{s1} & 1\\
	\end{bmatrix}\right)
	+
	(1-\pi_1)
	N\left(
	\begin{bmatrix}
	\mu_2\\
	\mu_2\\
	\end{bmatrix},
	\begin{bmatrix}
	1 & \theta_{s2}\\
	\theta_{s2} & 1\\
	\end{bmatrix}\right),  \quad\text{for $s=0,1$}
	\]
We also consider the cases that the reproducibility of the two workflows are indifferent ($\theta_{02}=\theta_{12}=0.6$) or different in strong candidates ($\theta_{02}=0.6$ and $\theta_{12}=0.9$). In both cases, we set all the other parameters identical between the workflows, with 
 $\mu_1=0$, $\mu_2=2.5$, and $\pi_1=0.6$, $0.8$, $0.9$, and $0.95$, as in Section \ref{SS:baseline}.  
For each case, we consider the situations that the scores of weak candidates are either uncorrelated ($\theta_{01}=0$) or correlated ($\theta_{01}=0.4$) across replicates. 
\end{itemize}

For each scenario,  we simulate 100 datasets,  each of which consists of $n=10,000$ pairs of observations.  We fit both scenarios with the model (\ref{mod4}), 
	\begin{align*}\label{modmixnorm0}
	\begin{aligned}
	\log\left(\Psi(t)|x\right) 
	&= \beta_{01} \left[\log(t)-\log(\tau)\right]_{-} + 
	\beta_{02} \left[\log(\tau) + \left\{\log(t)-\log(\tau)\right\}_{+}\right]\\
	&+ x\left(
	\beta_{11} \left[\log(t)-\log(\tau)\right]_{-} + 
	\beta_{12} \left[\log(\tau) + \left\{\log(t)-\log(\tau)\right\}_{+}\right]
	\right),
	\end{aligned}
	\end{align*}
	where $x=0, 1$ for workflow 0 and workflow 1, respectively. 
The regression coefficients $\beta_{11}$ and $\beta_{12}$ show the difference
in reproducibility between the workflows for candidates below and above the change-point $\tau$, respectively.   
We fit this model with $M=100$ equally spaced cutoffs in (0,1). The standard errors of the estimates  are obtained by the bootstrap method with 100 resampling times. The difference in reproducibility is tested using Wald's test for $H_0: \beta_{1j}=0$	(j=1, 2) at the significance level of 0.05. 

The results are summarized in Tables~\ref{tab:pw_gumnorm}---\ref{tab:pw_bivnorm1}.  
As shown in these tables, the estimated change-point $\tau$ accurately approximates the mixing proportion $\pi_1$ in all simulations.  
We can draw the conclusions as below:
	\begin{itemize}
	\item \textbf{Scenario~I with covariates}:
	When the reproducibility of the two workflows are similar,
	the empirical type I errors of $\beta_{11}$ and $\beta_{12}$ both are close to the nominal level of 0.05 for all mixing proportions.
	\item \textbf{Scenario~II with covariates}:
	When the two workflows have different levels of reproducibility for strong candidates, 
	  $\beta_{12}$ has a high power to detect the difference, especially when there is a sizable proportion of strong candidates ($\pi_1=0.6$ and $0.8$). The power is lower when the data is dominated by weak candidates ($\pi_1=0.9$ and $0.95$).  This is expected, because the information to distinguish the workflows is very limited in these cases. 
	  The type I error for $\beta_{11}$ is still reasonably controlled at the nominal level for most $\pi_1$'s, correctly reflecting the similarity in reproducibility between the two workflows for weak candidates. 
	 When $\pi_1=0.6$ or $0.8$, the type I error for $\beta_{11}$ is slightly elevated. 
	This is because the difference between workflows in the strong candidates ($\beta_{12}$) spills over to the weak ones ($\beta_{11}$).
	This effect is negligible when the proportion of relevant candidates is low.  Again, these results demonstrate that our method is robust even when there is a model misspecification.
\end{itemize}

\begin{table}[ht]
\centering
\caption{
Estimated segmented model for comparing the reproducibility of two workflows in Scenario~I. 
Same reproducibility: $\theta_{12}=\theta_{22}=1.2$; 
Different reproducibility: $\theta_{12}=1.2, \theta_{22}=2.0$. 
Est: average of the estimated value; SD: standard deviation of the estimated values; ESE: estimated standard error; Power: power for detecting difference in reproducibility. 
}
\label{tab:pw_gumnorm}
\scalebox{0.9}{
\begin{tabular}{rcrrrrr c rrrrr}
 \hline\hline
 & &\multicolumn{5}{c}{ Same reproducibility} & &  \multicolumn{5}{c}{Different reproducibility}\\
  & &\multicolumn{5}{c}{$(\theta_{12}=1.2$, $\theta_{22}=1.2)$} & &  \multicolumn{5}{c}{$(\theta_{12}=1.2$, $\theta_{22}=2.0)$}\\
\cline{3-7} \cline{9-13}\noalign{}
\multicolumn{1}{c}{$\pi_1$} &  & 
\multicolumn{1}{c}{$\widehat{\tau}$} & 
\multicolumn{1}{c}{$\widehat{\beta}_{01}$} & 
\multicolumn{1}{c}{$\widehat{\beta}_{02}$} & 
\multicolumn{1}{c}{$\widehat{\beta}_{11}$} & 
\multicolumn{1}{c}{$\widehat{\beta}_{12}$} & 
& 
\multicolumn{1}{c}{$\widehat{\tau}$} & 
\multicolumn{1}{c}{$\widehat{\beta}_{01}$} & 
\multicolumn{1}{c}{$\widehat{\beta}_{02}$} & 
\multicolumn{1}{c}{$\widehat{\beta}_{11}$} & 
\multicolumn{1}{c}{$\widehat{\beta}_{12}$}
\\ 
\cline{3-7} \cline{9-13}\noalign{}
 0.600 & Est & 0.559 & 1.911 & 1.260 & 0.002 & -0.000 &  & 0.564 & 1.904 & 1.256 & 0.012 & -0.045 \\ 
		& SD & 0.008 & 0.019 & 0.007 & 0.025 & 0.008 &  & 0.011 & 0.022 & 0.007 & 0.025 & 0.009 \\ 
 		& ESE & 0.006 & 0.020 & 0.007 & 0.026 & 0.008 &  & 0.009 & 0.022 & 0.007 & 0.026 & 0.009 \\ 
	    & Power &  &   &  & 0.030 & 0.030 &  &  &  &   & 0.080 & 0.990 \\ 
0.800 & Est & 0.773 & 1.937 & 1.288 & 0.002 & 0.000 & & 0.778 & 1.933 & 1.282 & 0.010 & -0.031 \\ 
	    & SD & 0.008 & 0.016 & 0.013 & 0.020 & 0.014 & & 0.007 & 0.015 & 0.010 & 0.020 & 0.014 \\ 
         & ESE & 0.008 & 0.016 & 0.014 & 0.021 & 0.014 & & 0.009 & 0.016 & 0.013 & 0.020 & 0.015 \\ 
        & Power &  &  &   & 0.060 & 0.040 &  &   &   &   & 0.090 & 0.580 \\ 
 0.900 & Est & 0.880 & 1.960 & 1.351 & 0.002 & 0.002 & & 0.886 & 1.955 & 1.337 & 0.008 & -0.031 \\ 
        & SD & 0.006 & 0.013 & 0.020 & 0.019 & 0.023 & & 0.010 & 0.015 & 0.025 & 0.019 & 0.025 \\ 
      & ESE & 0.007 & 0.014 & 0.021 & 0.018 & 0.022 &  & 0.009 & 0.015 & 0.022 & 0.018 & 0.023 \\ 
      & Power &   &   &   & 0.050 & 0.060 & &   &   &   & 0.060 & 0.330 \\ 
 0.950 & Est & 0.938 & 1.974 & 1.408 & 0.001 & 0.002 & & 0.940 & 1.972 & 1.402 & 0.005 & -0.030 \\ 
	    & SD & 0.002 & 0.013 & 0.022 & 0.018 & 0.028 &  & 0.005 & 0.013 & 0.030 & 0.018 & 0.031 \\ 
        & ESE & 0.004 & 0.013 & 0.028 & 0.017 & 0.031 & & 0.005 & 0.013 & 0.030 & 0.017 & 0.034 \\ 
	   & Power &   &   &   & 0.030 & 0.020 &   &  &   &   & 0.050 & 0.140 \\ 
   \hline
\end{tabular}
}
\end{table}

\begin{table}[ht]
\centering
\caption{
Estimated segmented model for comparing the reproducibility of two workflows in Scenario~II,
when the scores of reproducible candidates are uncorrelated across replicates 
(i.e., $\theta_{11}=0$). 
Estimated segmented model for comparing the reproducibility of two workflows in Scenario~I. 
Same reproducibility: $\theta_{12}=\theta_{22}=0.6$; 
Different reproducibility: $\theta_{12}=0.6, \theta_{22}=0.9$. 
Est: average of the estimated value; SD: standard deviation of the estimated values; ESE: estimated standard error; Power: power 
 for detecting difference in reproducibility.
}
\label{tab:pw_bivnorm0}
\scalebox{0.9}{
\begin{tabular}{rcrrrrr c rrrrr}
 \hline\hline
 & &\multicolumn{5}{c}{ Same reproducibility} & &  \multicolumn{5}{c}{Different reproducibility}\\
  & &\multicolumn{5}{c}{$(\theta_{12}=0.6$, $\theta_{22}=0.6)$} & &  \multicolumn{5}{c}{$(\theta_{12}=0.6$, $\theta_{22}=0.9)$}\\
\cline{3-7} \cline{9-13}\noalign{}
\multicolumn{1}{c}{$\pi_1$} &  & 
\multicolumn{1}{c}{$\widehat{\tau}$} & 
\multicolumn{1}{c}{$\widehat{\beta}_{01}$} & 
\multicolumn{1}{c}{$\widehat{\beta}_{02}$} & 
\multicolumn{1}{c}{$\widehat{\beta}_{11}$} & 
\multicolumn{1}{c}{$\widehat{\beta}_{12}$} & 
& 
\multicolumn{1}{c}{$\widehat{\tau}$} & 
\multicolumn{1}{c}{$\widehat{\beta}_{01}$} & 
\multicolumn{1}{c}{$\widehat{\beta}_{02}$} & 
\multicolumn{1}{c}{$\widehat{\beta}_{11}$} & 
\multicolumn{1}{c}{$\widehat{\beta}_{12}$}
\\ 
\cline{3-7} \cline{9-13}\noalign{}
 0.600 & Est & 0.563 & 1.869 & 1.295 & 0.002 & 0.001 &  & 0.580 & 1.851 & 1.286 & 0.028 & -0.069 \\ 
	     & SD & 0.010 & 0.021 & 0.008 & 0.026 & 0.007 &  & 0.012 & 0.023 & 0.008 & 0.025 & 0.007 \\ 
		& ESE & 0.011 & 0.021 & 0.008 & 0.026 & 0.009 &   & 0.013 & 0.023 & 0.008 & 0.025 & 0.010 \\ 
		& Power &   &   &   & 0.070 & 0.000 &   &   &   &  & 0.190 & 1.000 \\ 
 0.800 & Est & 0.778 & 1.916 & 1.350 & -0.001 & -0.001 &   & 0.789 & 1.908 & 1.338 & 0.013 & -0.062 \\ 
	    & SD & 0.012 & 0.020 & 0.015 & 0.021 & 0.012 &  & 0.011 & 0.021 & 0.013 & 0.021 & 0.012 \\ 
	  	& ESE & 0.012 & 0.017 & 0.016 & 0.021 & 0.015 &  & 0.012 & 0.017 & 0.015 & 0.020 & 0.017 \\ 
	     & Power &   &   &   & 0.080 & 0.020 & &   &   &   & 0.100 & 0.980 \\ 
 0.900 & Est & 0.885 & 1.945 & 1.422 & 0.002 & -0.003 & & 0.895 & 1.939 & 1.401 & 0.010 & -0.067 \\ 
	    & SD & 0.011 & 0.015 & 0.026 & 0.017 & 0.015 &   & 0.007 & 0.014 & 0.019 & 0.017 & 0.016 \\ 
         & ESE & 0.011 & 0.015 & 0.027 & 0.018 & 0.023 &   & 0.010 & 0.014 & 0.024 & 0.018 & 0.025 \\ 
	    & Power &   &   &  & 0.020 & 0.000 &   &   &   &   & 0.090 & 0.810 \\ 
0.950 & Est & 0.940 & 1.966 & 1.489 & 0.002 & -0.003 &   & 0.951 & 1.960 & 1.444 & 0.006 & -0.077 \\ 
	    & SD & 0.009 & 0.013 & 0.042 & 0.016 & 0.019 & & 0.010 & 0.014 & 0.047 & 0.015 & 0.029 \\ 
	    & ESE & 0.009 & 0.013 & 0.043 & 0.017 & 0.033 &   & 0.009 & 0.014 & 0.042 & 0.017 & 0.038 \\ 
	    & Power &   &   &   & 0.020 & 0.000 &  &   &  &   & 0.040 & 0.530 \\ 
   \hline
\end{tabular}
}
\end{table}

\begin{table}[ht]
\centering
\caption{Estimated segmented model for comparing the reproducbility of two workflows in Scenario~II,
when the scores of irreproducible candidates are correlated across replicates 
(i.e., $\theta_{11}=0.4$).  
Est: average of the estimated value; SD: standard deviation of the estimated values; ESE: estimated standard error; Power: power 
 for detecting difference in reproducibility.
}
\label{tab:pw_bivnorm1}
\scalebox{0.9}{
\begin{tabular}{rcrrrrr c rrrrr}
 \hline\hline
 & &\multicolumn{5}{c}{ Same reproducibility} & &  
 \multicolumn{5}{c}{Different reproducibility}\\
  & &\multicolumn{5}{c}{$(\theta_{12}=0.6$, $\theta_{22}=0.6)$} & &  \multicolumn{5}{c}{$(\theta_{12}=0.6$, $\theta_{22}=0.9)$}\\
\cline{3-7} \cline{9-13}\noalign{}
\multicolumn{1}{c}{$\pi_1$} &  & 
\multicolumn{1}{c}{$\widehat{\tau}$} & 
\multicolumn{1}{c}{$\widehat{\beta}_{01}$} & 
\multicolumn{1}{c}{$\widehat{\beta}_{02}$} & 
\multicolumn{1}{c}{$\widehat{\beta}_{11}$} & 
\multicolumn{1}{c}{$\widehat{\beta}_{12}$} & 
& 
\multicolumn{1}{c}{$\widehat{\tau}$} & 
\multicolumn{1}{c}{$\widehat{\beta}_{01}$} & 
\multicolumn{1}{c}{$\widehat{\beta}_{02}$} & 
\multicolumn{1}{c}{$\widehat{\beta}_{11}$} & 
\multicolumn{1}{c}{$\widehat{\beta}_{12}$}
\\ 
\cline{3-7} \cline{9-13}\noalign{}
 0.60 & Est & 0.561 & 1.522 & 1.259 & 0.000 & 0.001 &&
  0.580 & 1.513 & 1.256 & 0.024 & -0.068\\ 
  & SD & 0.010 & 0.013 & 0.006 & 0.016 & 0.008 &&
   0.015 & 0.014 & 0.005 & 0.015 & 0.009\\ 
  & ESE & 0.010 & 0.014 & 0.006 & 0.018 & 0.008 &&
   0.017 & 0.016 & 0.006 & 0.018 & 0.010\\ 
 & Power &  &  &  & 0.010 & 0.040 &&
   &  &  & 0.250 & 1.000\\   
 0.80 & Est & 0.786 & 1.561 & 1.304 & 0.000 & -0.001 &&
 0.806 & 1.554 & 1.294 & 0.015 & -0.072\\ 
 & SD & 0.013 & 0.011 & 0.010 & 0.014 & 0.011 &&
 0.012 & 0.011 & 0.009 & 0.013 & 0.013\\ 
 & ESE & 0.014 & 0.011 & 0.012 & 0.015 & 0.013 &&
 0.014 & 0.011 & 0.010 & 0.015 & 0.016\\ 
 & Power &  &  &  & 0.040 & 0.030 &&
   &  &  & 0.160 & 1.000\\   
0.90 & Est & 0.901 & 1.589 & 1.355 & -0.001 & -0.001 && 
0.912 & 1.586 & 1.344 & 0.008 & -0.085\\ 
 & SD & 0.010 & 0.009 & 0.016 & 0.012 & 0.016 &&
 0.009 & 0.009 & 0.014 & 0.012 & 0.020\\ 
 & ESE & 0.012 & 0.010 & 0.019 & 0.013 & 0.022 &&
 0.010 & 0.010 & 0.017 & 0.013 & 0.026\\ 
 & Power &  &  &  & 0.070 & 0.000 &&
  &  &  & 0.070 & 0.910\\    
0.95 & Est & 0.959 & 1.608 & 1.390 & -0.002 & -0.002 &&
0.963 & 1.607 & 1.379 & 0.003 & -0.108\\ 
 & SD & 0.005 & 0.008 & 0.026 & 0.012 & 0.031 &&
  0.009 & 0.009 & 0.028 & 0.012 & 0.049\\ 
 & ESE & 0.007 & 0.009 & 0.030 & 0.013 & 0.036 &&
 0.008 & 0.009 & 0.031 & 0.012 & 0.048 \\ 
 & Power &  &  &  & 0.080 & 0.020 &&
   &  &  & 0.040 & 0.650 \\    
   \hline
\end{tabular}
}
\end{table}

\subsection{Performance of the test for the existence of a change point.} 

We use Scenario I in Section \ref{SS:baseline} to assess the performance of the test procedure in Section~2.5. 
Because the proportion of weak candidates is usually large in practice,
we evaluate the performance mainly in this range, at the mixing proportion of the weak candidates of 
$\pi_1$ = 0.0, 0.80, 0.90, 
0.95, 0.96, 0.97, 0.98, 0.99, and 1.0.  
When $\pi_1=0.0$ or 1.0, all the candidates are from a single component, either all reproducible or all irreproducible. 

For each parameter settings,  we simulate 100 datasets with $n=10,000$ pairs of observations 
and obtain p-values with 100 bootstrap replicates with the nominal significance level of $0.05$.   
As shown in Table \ref{tab:test},  when the candidate list consists of a single population   ($\pi_1=0.0$ or 1.0), the power is nearly zero, consistent with the homogeneity in the data.  
When the  proportion of weak candidates is between 0.80 and 0.96, 
the test detects the heterogeneity of reproducibility with a power of nearly 1. 
However,  when $\pi_1$ is  between 0.97 and 0.99,  the power is very low. 
This is not surprising, as the candidate lists are dominated by a single population in these cases, making it very difficult to detect the heterogeneity. 
In short, this simulation confirms that  the test is valid and effective.

\begin{table}[ht]
\centering
\caption{Power for testing the existence of a change point in Scenario I baseline model.}\label{tab:test}
\begin{tabular}{crrrrrrrrr}
  \hline\hline
  $\pi_1$& 0.00 & 0.80 & 0.90 &  0.95 & 0.96 & 0.97 & 0.98 & 0.99 & 1.00\\ 
  \hline
Power & 0.00 &  1.00&1.00 & 1.00 &0.94& 0.29& 0.01 & 0.01 & 0.01\\
   \hline
\end{tabular}
\end{table}

\section{Applications.}\label{sec:applications}

\subsection{ChIP-seq study}

We apply our methods to the Su(HW) ChIP-seq data in \citet{chen2012systematic} to characterize the effect of sequencing depth on the reproducibility of binding site identification.  
Among all peak calling algorithms in their analyses, Useq is consistently a top performer in all aspects of their evaluation. 
Therefore, we use the binding sites identified by the peak calling algorithm Useq for our analysis.
Similar to \citet{chen2012systematic}, we focus on the binding sites that are commonly identified on both replicates. At the sequencing depth of 0.45, 0.9, 2.7, 5.4 and 16.2 million reads,  1,335, 2,198, 3,813, 4,631, and 5,499 binding sites are identified on both replicates, respectively. 
We compute $\Psi(t)$ based on the scores assigned by Useq to the two biological replicates, using $M=100$ equally spaced cutoffs in $(0,1)$. 
The rank scatterplots of the scores on the two biological replicates suggest that the Gumble-Hougaard functional form in \eqref{mod_aim} is a reasonable fit. So
we fit the model 
\[
   \log\left(\Psi(t)\right) 
     =  
    \sum_{s=0}^{4}\left(\beta_{s1}\left\{\log(t)-\log(\tau)\right\}_- + 
    \beta_{s2}\left[\log(\tau) +\left\{\log(t)-\log(\tau)\right\}_+\right] \right)X_s,
\]
where 0.45M is used as the baseline, $X_0=1$ represents baseline, and $X_1 \ldots, X_4$ are the dummy variables for the depths of 0.9M, 2.7M, 5.4M and 16.2M, respectively.

\begin{figure}[ht]
  \centering	
	\hspace{1em}
	\begin{subfigure}[t]{0.45\textwidth}
		\includegraphics[height=0.25\textheight, width=\textwidth]{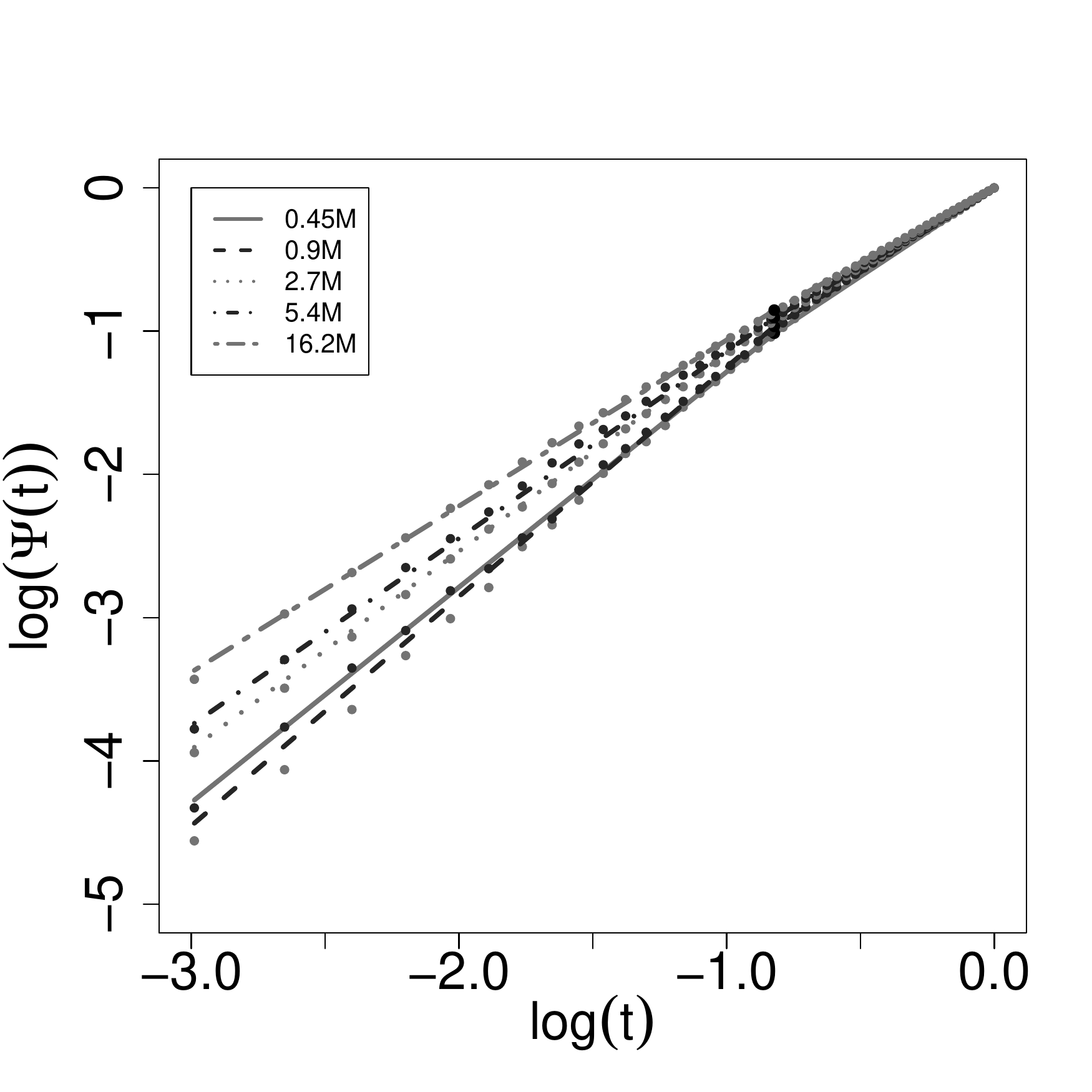}
		\caption{}
	\end{subfigure}
	\begin{subfigure}[t]{0.45\textwidth}
		\includegraphics[height=0.25\textheight, width=\textwidth]{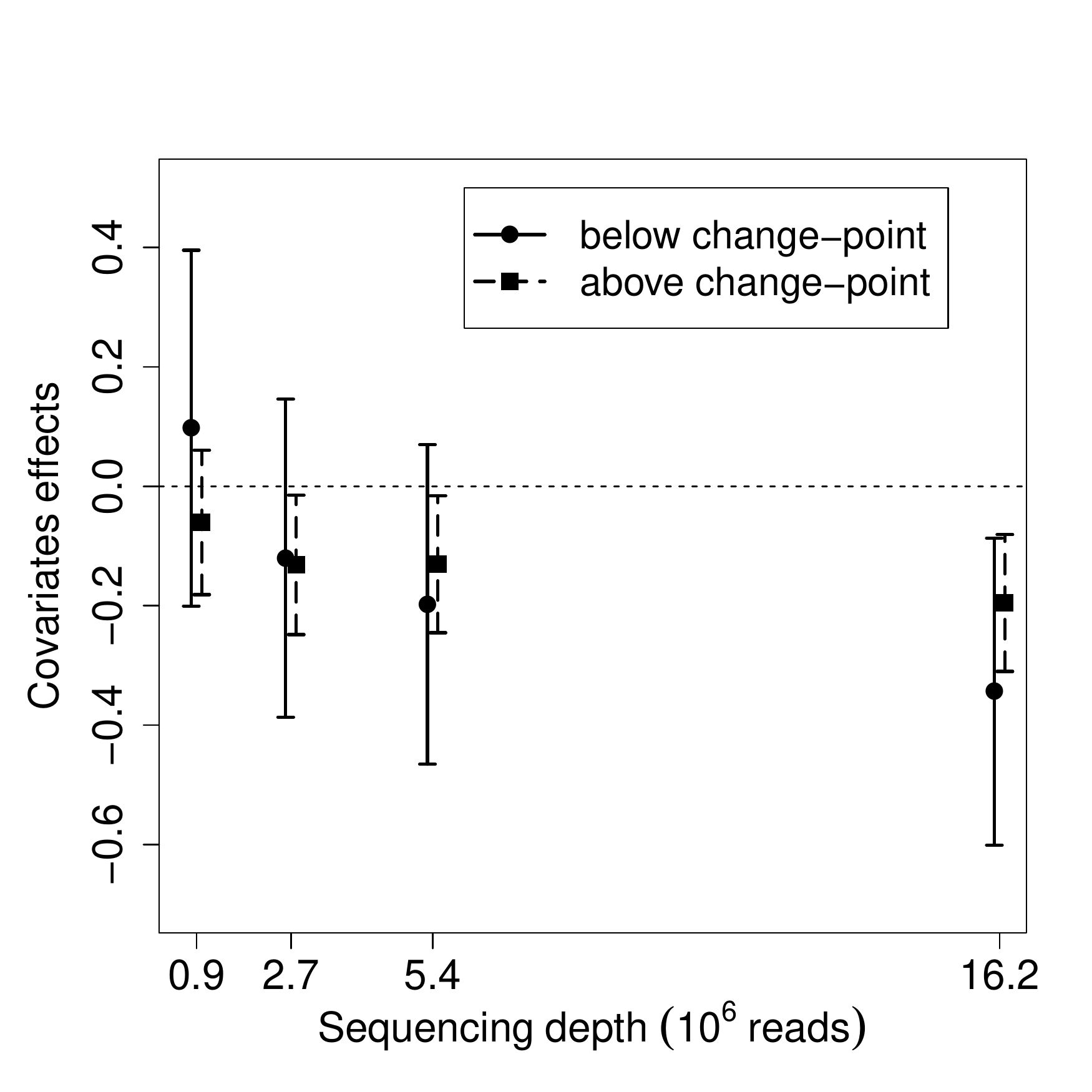}
		\caption{}
	\end{subfigure}
     \caption{
     	Estimated segmented regression model for the ChIP-seq data.
     	(a) Relationship between $\log(t)$ and $\log(\Psi(t))$ at different sequencing depths and the fitted line. Dots: empirical data; Lines: fitted lines.
     	(b) The estimated covariate effects at different sequencing depths in irreproducible ($\beta_{s1}$) and reproducible ($\beta_{s2}$) peaks. The dotted line marks zero covariate effect. The 95\% confidence intervals are shown. 
     }
     \label{fig:ChIP-results}
\end{figure}

The estimated parameters are summarized in Table~\ref{tab:ChIP}.  
The estimated change point is 0.440, indicating that the impact of sequencing depth on the top half and the bottom half of the putative binding sites is different. 
Though both groups show increased reproducibility with the increase of sequencing depth (Figure~\ref{fig:ChIP-results}), the improvement pattern is different. 
For strong peaks, the initial increase of sequencing depth results in a fast improvement of reproducibility. After the sequencing depth reaches 2.7M, the reproducibility is significantly improved  (95\% CI for $\hat{\beta}_{22}$: $[-0.248, -0.015]$) over 0.45M. 
Further increase of sequencing depth leads to very limited improvement, with overlapping confidence intervals for the coefficients ($\hat{\beta}_{s2}$ in Table~\ref{tab:ChIP}). For example, for the highest ranked $20\%$ peaks (i.e. $t=0.80$), the estimated probabilities of being reproducible are 
$t^{\hat{\beta}_{12}}= 1.013$ ($95\%$ CI: [0.987, 1.041]), 
$t^{\hat{\beta}_{22}}=1.030$ ($95\%$ CI: [1.003, 1.057]), 
$t^{\hat{\beta}_{32}}=1.030$ ($95\%$ CI: [1.004, 1.056]), 
and $t^{\hat{\beta}_{42}}=1.045$ ($95\%$ CI: [1.018, 1.072]) 
times as high at 0.9M, 2.7M,  5.4M, and 16.2M as that at 0.45M, respectively. 
However, for weak peaks, the reproducibility does not show significant improvement until the sequencing depth reaches 16.2M (95\% CI for $\hat{\beta}_{41}$:$[-0.601, -0.087]$).   For example, the estimated probabilities of being reproducibly reported in the bottom $20\%$ (i.e. $t=0.20$) are 
$t^{\hat{\beta}_{11}}= 0.855$ ($95\%$ CI: [0.530, 1.382]), 
$t^{\hat{\beta}_{21}}=1.213$ ($95\%$ CI: [0.791, 1.864]), 
$t^{\hat{\beta}_{31}}=1.375$ ($95\%$ CI: [0.893, 2.117]), 
and $t^{\hat{\beta}_{41}}=1.740$ ($95\%$ CI: [1.150, 2.631]) 
times as high at 0.9M, 2.7M,  5.4M, and 16.2M as that at 0.45M, respectively. 

Together, this indicates that 2.7M probably is sufficient to produce reproducible identifications for major binding sites. Further increase of sequencing depth has limited return. However, if one is interested to categorize most binding sites, then 16.2M is likely necessary.     

\begin{table}[ht]
	\caption{Sequence depth effects on the reproducibility of binding sites identified using the Useq algorithm from a ChIP-seq experiment 
	in \citet{chen2012systematic}.
	}\label{tab:ChIP}
	\centering
	(a) Data structure for ChIP-seq experiment study.
	\begin{tabular}{ccccccccccccccc}
		\hline\hline
		Depth &\multicolumn{2}{c}{0.45M} && \multicolumn{2}{c}{0.9M} && \multicolumn{2}{c}{2.7M} && \multicolumn{2}{c}{5.4M} && \multicolumn{2}{c}{16.2M}\\ 
		\cline{2-3}\noalign{}  \cline{5-6}\noalign{} \cline{8-9}\noalign{} \cline{11-12}\noalign{} \cline{14-15}\noalign{}
		  & rep1 & rep2 && rep1 & rep2 && rep1 & rep2 && rep1 & rep2 && rep1 & rep2\\ \hline
		  & $y_{11}^0$ & $y_{12}^0$ && $y_{11}^1$ & $y_{12}^1$ && $y_{11}^2$ & $y_{12}^2$ && $y_{11}^3$ & $y_{12}^3$ && $y_{11}^4$ & $y_{12}^4$\\  
	      & $\vdots$& $\vdots$&& $\vdots$ & $\vdots$&& $\vdots$ & $\vdots$ && $\vdots$  &  $\vdots$  && $\vdots$  &  $\vdots$   \\
		  & $y_{n_0,1}^0$ & $y_{n_0,2}^0$ && $y_{n_1,1}^1$ & $y_{n_1,2}^1$ && $y_{n_2,1}^2$ & $y_{n_2,2}^2$ && $y_{n_3,1}^3$ & $y_{n_3,1}^3$ && $y_{n_4,1}^4$ & $y_{n_4,2}^4$ \\ \hline
	\end{tabular}
		
\vspace{10pt}
(b) Estimated change point and sequencing depth effects. The  baseline is 0.45M. \\
\begin{tabular}{rcrc}
	\hline\hline
 \multicolumn{1}{c}{} &\multicolumn{1}{c}{} & \multicolumn{1}{c}{Estimate} &   \multicolumn{1}{c}{$95\%$ confidence interval}  \\ 
	\hline
	Change point        & $\tau$ & 0.440 &   [0.352, 0.527] \\ 
	Baseline & $\beta_{01}$ & 1.503 & [1.246, 1.761] \\ 
	      & $\beta_{02}$  & 1.237 & [1.125, 1.350] \\ 
     0.9M    & $\beta_{11}$  & 0.097 & [-0.201,0.395] \\ 
                 & $\beta_{12}$  &-0.060 &  [-0.181, 0.060] \\ 
     2.7M   & $\beta_{21}$  &-0.120 & [-0.387, 0.146] \\ 
              & $\beta_{22}$  &-0.132 &  [-0.248, -0.015] \\ 
	   5.4M  & $\beta_{31}$  &-0.198 &  [-0.466, 0.070] \\ 
		  & $\beta_{32}$  &-0.131 &  [-0.245, -0.016] \\ 
		16.2M   & $\beta_{41}$  &-0.344 &  [-0.601, -0.087] \\ 
		      & $\beta_{42}$  &-0.196 &  [-0.310, -0.081] \\ 
		\hline
	\end{tabular}

\end{table}

\subsection{Salmon data}
In addition to assessing reproducibility, our method can be also used to explore the change of association between a pair of variables in other contexts. 
Here we illustrate this use using a dataset from \citet[][p.161]{simonoff1996smoothing}. 
This dataset studies the relationship between the size of the annual spawning stock (in thousands of fish) and its production of new catchable-sized fish (recruits) for the Skeena River sockeye salmon stock from 1940 to 1967. 
The scatterplot (Figure~\ref{fig:salmon}a) shows that the less numerous spawning stocks have a weaker association with the production of recruits, whereas the more numerous spawning stocks have a stronger association.  
\citet{fisher2001graphical} speculates that the heterogeneous association indicates that the data may contain a mixed fish populations.
Using the derivative of the correspondence curve, \citet{li2011measuring} visually identifies a change of association in the data. 
 However, the existence and location of transition have not been formally quantified. Notably, this dataset consists of only 28 measurements, thus it is illustrative of the small sample behavior of our method.

Here we test the existence of a transition and estimate the location of a change point using our method.  Specifically,  we fit the model 
\begin{align*}
\log\left(\Psi(t)\right) 
=
 \alpha_1\left(\log(t)-\log(\tau)\right)_- + \alpha_2\left[\log(\tau) + \left(\log(t)-\log(\tau)\right)_+ \right].
\end{align*} 
Since there are only 28 points, 
we compute $\Psi(t)$  from the pairs of spawners and recruits using $M=28$ equally spaced cutoffs in $(0,1)$. 
We test the existence of a change point with 1000 bootstrap times, using the test proposed in Section 2.5. 
The p-value is nearly 0, indicating a highly significant heterogeneity in the association between spawning stock and production of recruits. 
The estimates  are
$\widehat{\alpha}_1= 1.102$ (SE = 0.409), 
 $\widehat{\alpha}_2= 1.753$ (SE =0.320),  
 with the change point $\widehat{\tau}=0.629$ (SE = 0.131).  
 This suggests that the top $37.1\%$ spawning stocks are from a strongly associated group and the bottom $62.9\%$ are from a weakly associated group.  
 
As a comparison, we also fit the data using the homogeneous model,  
$
	\log\left(\Psi(t)\right) 
=
 \beta_1\log(t),
$
 and obtain $\widehat{\beta}_1=1.400$ (SE = 0.265). 
 The fitted curves of the proposed model shows a much better fit than that of the homogeneous model (Figure~\ref{fig:salmon}b-c).  
To quantify the performance of model fitting, 
we compare the mean integrated squared errors (MISE) for the fitted curves as in Section~\ref{SS:baseline}. 
The MISE of the proposed model (0.0011) is 42.1\% less than that of the homogeneous model
(0.0019), showing a much improved fit. This demonstrates that our method performs well even when sample size is very small.

\begin{figure}
		\centering\label{fig:salmon}
			\begin{subfigure}[t]{0.3\textwidth}
			    \includegraphics[height=0.25\textheight, width=\textwidth]{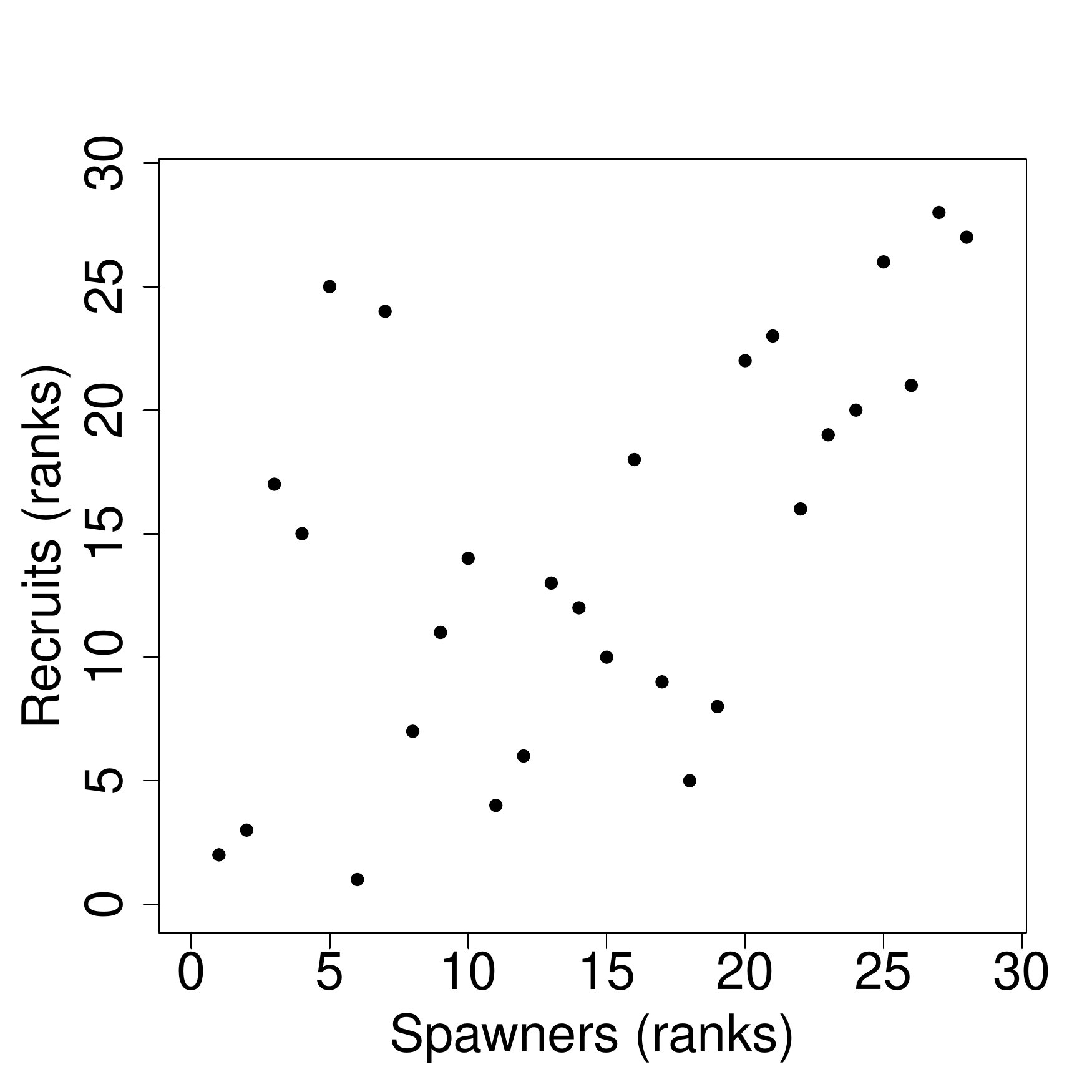}
			    \caption{}
			\end{subfigure}
			\hspace{-1em}
			\begin{subfigure}[t]{0.3\textwidth}
				\includegraphics[height=0.25\textheight, width=\textwidth]{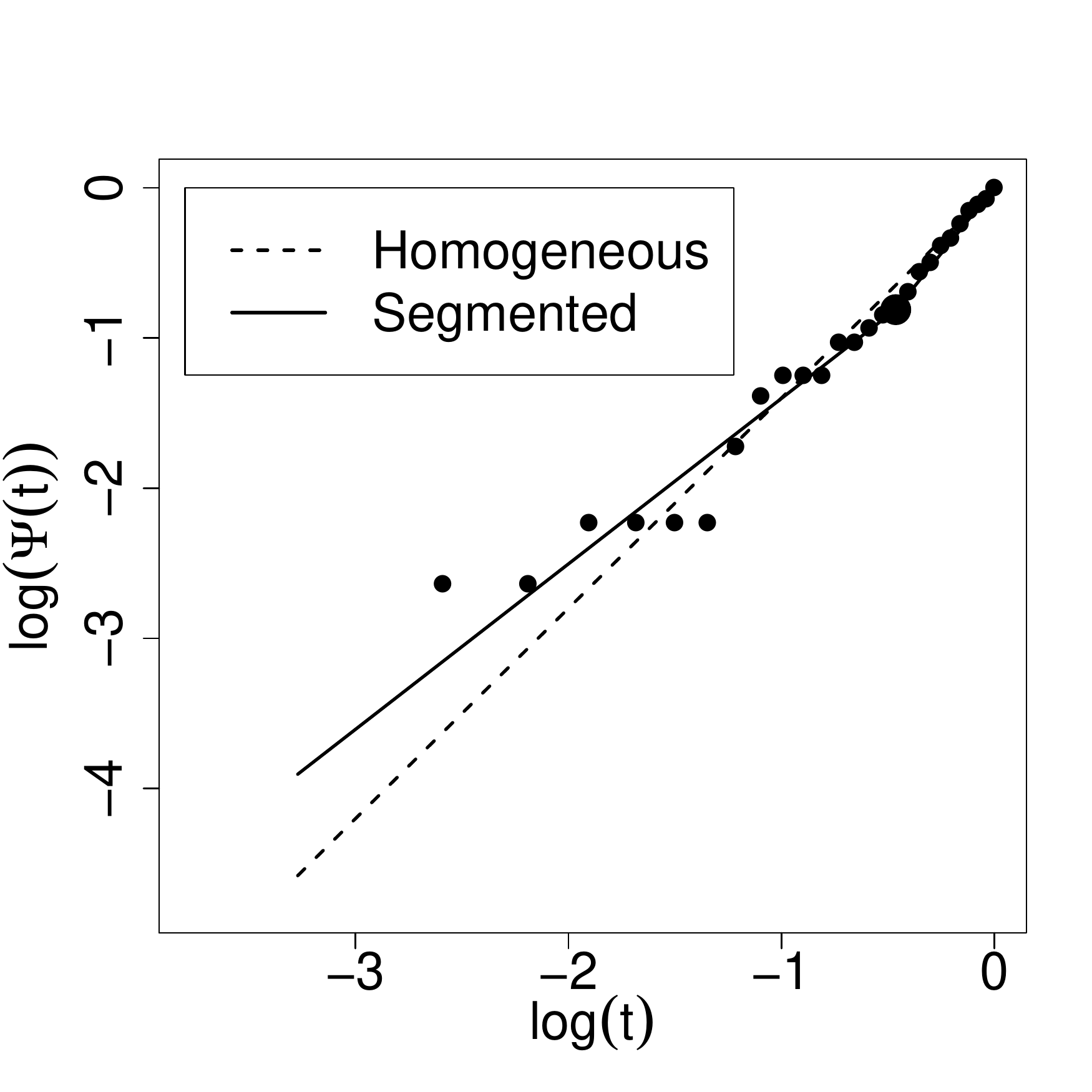}
				\caption{}
			\end{subfigure}
			\hspace{-1em}
			\begin{subfigure}[t]{0.3\textwidth}
 			    \includegraphics[height=0.25\textheight, width=\textwidth]{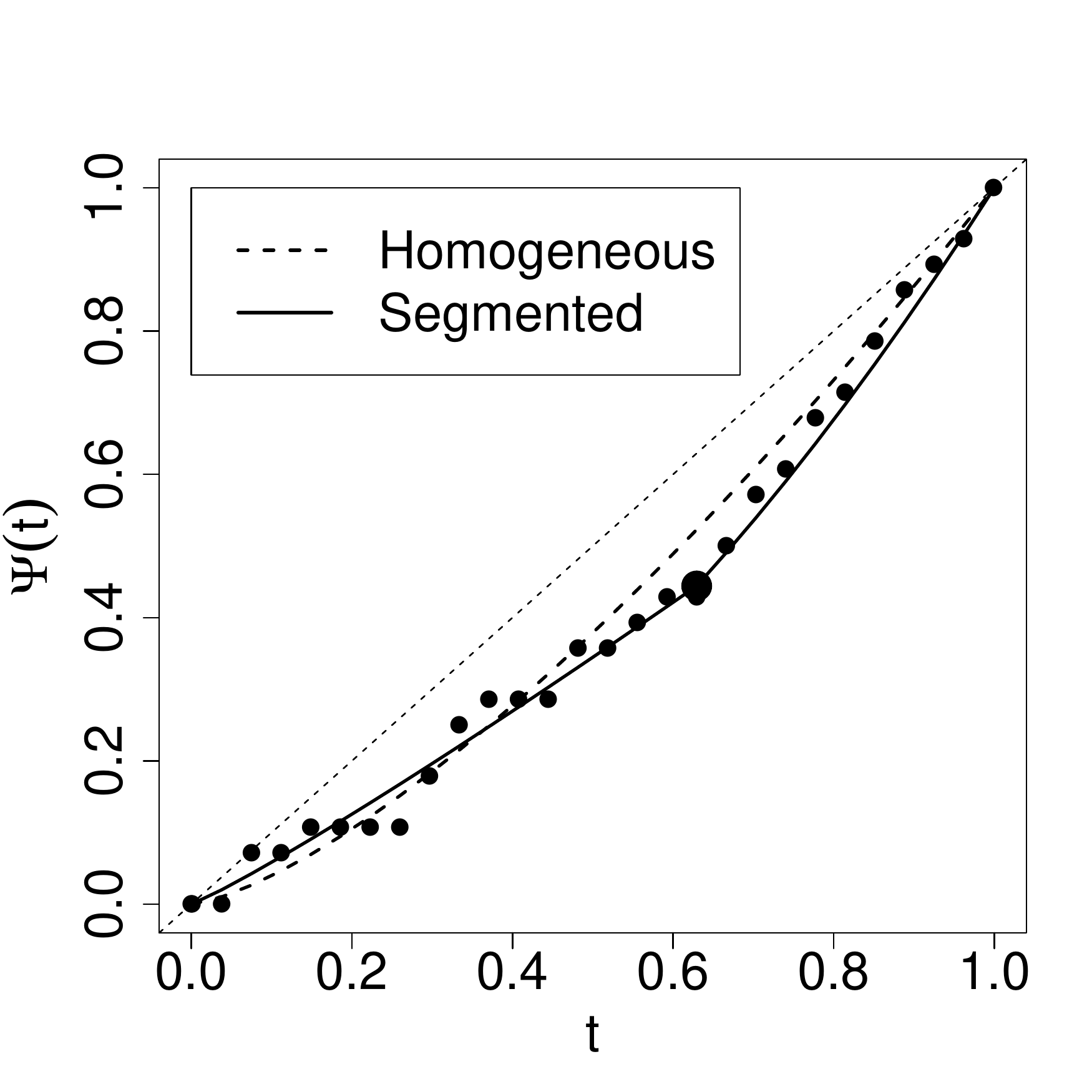}
 			    \caption{}
			\end{subfigure}
			\caption{Salmon data. (a) Rank scatterplot of the data. 
			(b) Exploratory data analysis shows a segmented linear trend between $\log(t)$ and $\log(\Psi(t))$. 
			(c) The fitted correspondence curve. 
		}
			\label{fig:salmon}
\end{figure}

\section{Conclusion.}
\label{sec:conclusion}

In this paper, 
we propose the segmented correspondence curve regression model for assessing the influence of operational factors on the reproducibility of ranked candidate lists obtained from high-throughput experiments. By extending the correspondence curve regression model to a segmented regression framework, it effectively captures the heterogeneity of reproducibility between strong and weak candidates, 
providing a detailed characterization of the effects of operational factors on both types of candidates.
Our results show that the proposed estimation procedure has both desirable asymptotic properties and good small-sample performance. 
A quasi-likelihood ratio testing method is also provided to test the existence of a structural change in reproducibility and determine when the segmented regression approach is appropriate.

The application of our method to a ChIP-seq experiment 
 reveals interesting insights on how sequencing depth affects the reproducibility of identified binding sites and how this effect differs between binding sites at different levels of prominence. 
The regression coefficients estimated from our model provide an easy-to-use and statistically sound guideline for determining the minimum sequencing depth required for producing reproducible results.  This application demonstrates the usefulness of our method for guiding the design of reliable and cost-effective high-throughput workflows. 

While our method was developed for assessing reproducibility of high-throughput experiments, our demonstration on the salmon data shows that it can also be used to study the structural changes in the dependence structures of two variables in other areas,
 for example, comovement patterns between stock prices. Although a vast amount of literature exists for modeling structural changes in the mean structure, methods for studying changes in dependence  structures are much less developed.  This work contributes a useful addition to this area. 


Currently, our work focuses on the dependence structure of ranked candidate lists from replicated high-throughput experiments.  
However, this segmented framework can be readily expanded to other dependence structures. 
 \citet{li2017regression} has identified a series of functional forms of the correspondence curve regression for other dependence structures. 
An extension to these regression forms will be useful for expanding this framework to a broad class of applications. 

\bigskip
\begin{center}
{\large\bf SUPPLEMENTARY MATERIAL}
\end{center}

\begin{description}

	\item[cccr20180520-supp]Technical proofs. (.pdf file) 
	
	\item[segCCR-0.0.tar] R-package. The package also contains all datasets used as examples in the article. (GNU zipped tar file)]
	
\end{description}

%
%
%
%

\bibliographystyle{apalike}

\begin{thebibliography}{}

\bibitem[Andrews, 1993]{andrews1993tests}
Andrews, D. (1993).
\newblock Tests for parameter instability and structural change with unknown
  change point.
\newblock {\em Econometrica}, 61:821--856.

\bibitem[Bai, 1996]{bai1996testing}
Bai, J. (1996).
\newblock Testing for parameter constancy in linear regressions: an empirical
  distribution function approach.
\newblock {\em Econometrica}, 64:597--622.

\bibitem[Chan, 1993]{chan1993consistency}
Chan, K. (1993).
\newblock Consistency and limiting distribution of the least squares estimator
  of a threshold autoregressive model.
\newblock {\em Annals of Statistics}, 21:520--533.

\bibitem[Chappell, 1989]{chappell1989fitting}
Chappell, R. (1989).
\newblock Fitting bent lines to data, with applications to allometry.
\newblock {\em Journal of Theoretical Biology}, 138:235--256.

\bibitem[Chen et~al., 2012]{chen2012systematic}
Chen, Y., Negre, N., Li, Q., Mieczkowska, J.~O., Slattery, M., Liu, T., Zhang,
  Y., Kim, T.-K., He, H.~H., Zieba, J., et~al. (2012).
\newblock Systematic evaluation of factors influencing {ChIP}-seq fidelity.
\newblock {\em Nature Methods}, 9:609--614.

\bibitem[Chiu et~al., 2006]{chiu2006bent}
Chiu, G., Lockhart, R., and Routledge, R. (2006).
\newblock Bent-cable regression theory and applications.
\newblock {\em Journal of the American Statistical Association}, 101:542--553.

\bibitem[Cho and White, 2007]{cho2007testing}
Cho, J.~S. and White, H. (2007).
\newblock Testing for regime switching.
\newblock {\em Econometrica}, 75:1671--1720.

\bibitem[Feder, 1975]{feder1975asymptotic}
Feder, P.~I. (1975).
\newblock On asymptotic distribution theory in segmented regression
  problems-identified case.
\newblock {\em The Annals of Statistics}, 3:49--83.

\bibitem[Fisher and Switzer, 2001]{fisher2001graphical}
Fisher, N. and Switzer, P. (2001).
\newblock Graphical assessment of dependence: Is a picture worth 100 tests?
\newblock {\em The American Statistician}, 55:233--239.

\bibitem[Hansen, 1996]{hansen1996inference}
Hansen, B.~E. (1996).
\newblock Inference when a nuisance parameter is not identified under the null
  hypothesis.
\newblock {\em Econometrica}, 64:413--430.

\bibitem[Hansen, 2017]{hansen2017regression}
Hansen, B.~E. (2017).
\newblock Regression kink with an unknown threshold.
\newblock {\em Journal of Business \& Economic Statistics}, 35:228--240.

\bibitem[Hinkley, 1969]{hinkley1969inference}
Hinkley, D.~V. (1969).
\newblock Inference about the intersection in two-phase regression.
\newblock {\em Biometrika}, 56:495--504.

\bibitem[Irizarry et~al., 2005]{irizarry2005multiple}
Irizarry, R.~A., Warren, D., Spencer, F., Kim, I.~F., Biswal, S., Frank, B.~C.,
  Gabrielson, E., Garcia, J.~G., Geoghegan, J., Germino, G., et~al. (2005).
\newblock Multiple-laboratory comparison of microarray platforms.
\newblock {\em Nature Methods}, 2:345.

\bibitem[Jung et~al., 2014]{jung2014impact}
Jung, Y.~L., Luquette, L.~J., Ho, J.~W., Ferrari, F., Tolstorukov, M., Minoda,
  A., Issner, R., Epstein, C.~B., Karpen, G.~H., Kuroda, M.~I., et~al. (2014).
\newblock Impact of sequencing depth in chip-seq experiments.
\newblock {\em Nucleic Acids Research}, 42:e74--e74.

\bibitem[Kosorok and Song, 2007]{kosorok2007inference}
Kosorok, M.~R. and Song, R. (2007).
\newblock Inference under right censoring for transformation models with a
  change-point based on a covariate threshold.
\newblock {\em The Annals of Statistics}, 35:957--989.

\bibitem[Landt et~al., 2012]{landt2012chip}
Landt, S.~G., Marinov, G.~K., Kundaje, A., Kheradpour, P., Pauli, F.,
  Batzoglou, S., Bernstein, B.~E., Bickel, P., Brown, J.~B., Cayting, P.,
  et~al. (2012).
\newblock Chip-seq guidelines and practices of the encode and modencode
  consortia.
\newblock {\em Genome Research}, 22:1813--1831.

\bibitem[Lee et~al., 2011]{lee2011testing}
Lee, S., Seo, M.~H., and Shin, Y. (2011).
\newblock Testing for threshold effects in regression models.
\newblock {\em Journal of the American Statistical Association}, 106:220--231.

\bibitem[{Li} et~al., 2011]{li2011measuring}
{Li}, Q., Brown, J.~B., Huang, H., and Bickel, P.~J. (2011).
\newblock Measuring reproducibility of high-throughput experiments.
\newblock {\em The Annals of Applied Statistics}, 5:1752--1779.

\bibitem[Li and Zhang, 2017]{li2017regression}
Li, Q. and Zhang, F. (2017).
\newblock A regression framework for assessing covariate effects on the
  reproducibility of high-throughput experiments.
\newblock {\em Biometrics}, https://doi.org/10.1111/biom.12832.

\bibitem[{Li, C.} et~al., 2011]{lich2011bent}
{Li, C.}, Wei, Y., Chappell, R., and He, X. (2011).
\newblock Bent line quantile regression with application to an allometric study
  of land mammals' speed and mass.
\newblock {\em Biometrics}, 67:242--249.

\bibitem[Nelsen, 2006]{nelsen2006introduction}
Nelsen, R.~B. (2006).
\newblock {\em An Introduction to Copulas}.
\newblock Springer.

\bibitem[Philtron et~al., 2017]{philtron2017maximum}
Philtron, D., Lyu, Y., Li, Q., and Ghosh, D. (2017).
\newblock Maximum rank reproducibility: A non-parametric approach to assessing
  reproducibility in replicate experiments.
\newblock {\em Journal of the American Statistical Association},
  https://doi.org/10.1080/01621459.2017.1397521.

\bibitem[Qu, 2008]{qu2008testing}
Qu, Z. (2008).
\newblock Testing for structural change in regression quantiles.
\newblock {\em Journal of Econometrics}, 146:170--184.

\bibitem[Quandt, 1958]{quandt1958estimation}
Quandt, R.~E. (1958).
\newblock The estimation of the parameters of a linear regression system
  obeying two separate regimes.
\newblock {\em Journal of the American Statistical Association}, 53:873--880.

\bibitem[Simonoff, 1996]{simonoff1996smoothing}
Simonoff, J.~S. (1996).
\newblock {\em Smoothing Methods in Statistics}.
\newblock Springer.

\bibitem[Zhang and Li, 2017a]{zhang2017threshold}
Zhang, F. and Li, Q. (2017a).
\newblock A continuous threshold expectile model.
\newblock {\em Computational Statistics \& Data Analysis}, 116:49--66.

\bibitem[Zhang and Li, 2017b]{zhang2017robust}
Zhang, F. and Li, Q. (2017b).
\newblock Robust bent line regression.
\newblock {\em Journal of Statistical Planning and Inference}, 185:41--55.

\end{thebibliography}

\end{document}


\maketitle

\section*{Appendix A.   Proofs.}

\underline{Regularity Conditions}. 
\begin{description}
	\item[(A1)] 
	$\tau_0=\arg\min\limits_{t\in \mathcal{T}} \ell\left(\widehat{\bbeta}(\tau), \tau \right)$  
	is unique,  where $\mathcal{T}$ is a compact set in $[0, 1]$.  
	
	\item[(A2)] 
	$\btheta$ is in $\Theta$, which is a compact subset of $\mathbb{R}^{p+3}$. 
	
	 \item[(A3)]  
	 Given $\beta_{s1}\neq \beta_{s2}$ for $s=0,1,\ldots, S$, the Hessian matrix  $A(\btheta_0)$ is nonsingular.  
	 
	 \item[(A4)] 
	 The covariate vector $\bx$ is bounded.  
\end{description}
Condition (A1) is the identifiability condition of the estimation.  
Conditions (A2)--(A4) are for the consistency and  the asymptotic normality of the estimates. 

\medskip 
\noindent
\textit{\underline {Proof of Theorem~1 
}}.  
We divide the proof in the following two steps. 

(i) We first establish the consistency of the estimators. 
We need to show that 
$$
	\sup_{\btheta\in \Theta}\left|\ell_{n}(\btheta)-\ell(\btheta)\right|  \mathop{\longrightarrow}^{P}0, 
$$ 
as n goes to infinity. 
To show that the class of functions $\{m_{\btheta}: \btheta\in \Theta\}$ is Glivenko-Cantelli, 
it is sufficient to show $m_{\btheta}$ is Lipschitz continuous,  
by the Glivenko-Cantelli theorem and Example 19.8 in \citet{van1998asymptotic}.   
Recall that $\btheta=(\bbeta^\top, t)^\top$ and the derivatives 
\begin{align*}
\begin{aligned}
	\frac{\partial m_{\btheta}}{\partial\bbeta} 
	&= 
	\mathop{\sum}\limits_{m=1}^M \mathop{\sum}\limits_{s=0}^S
	U_{im}
	\frac{ 
		\Delta \left(\exp\left\{\mathop{\sum}\limits_{s=0}^S x_s\bbeta_s^\top W(t_m, \tau)\right\}\bx\otimes W(t_m, \tau) \right)
	}
	{\Delta \left(\exp\left\{\mathop{\sum}\limits_{s=0}^S x_s\bbeta_s^\top W(t_m, \tau)\right\} \right)
	}\\
    \frac{\partial m_{\btheta}}{\partial \tau}
      &=  
     \mathop{\sum}\limits_{m=1}^M \mathop{\sum}\limits_{s=0}^S
     U_{im}
     \frac{
     	\Delta \left(\exp\left\{\mathop{\sum}\limits_{s=0}^S x_s\bbeta_s^\top W(t_m, \tau) \right\}
     	\mathop{\sum}\limits_{s=0}^S x_s(\beta_{s2}-\beta_{s1})\tau^{-1} I(t_m\leq \tau)  \right)
     }
     {
     	\Delta  \left(\exp\left\{\mathop{\sum}\limits_{s=0}^S x_s\bbeta_s^\top W(t_m, \tau) \right\} \right)
     }
\end{aligned}
\end{align*}

By the Condition (A2) and (A4),  
both the two terms 
$$
\frac{ 
	\Delta \left(\exp\left\{\mathop{\sum}\limits_{s=0}^S x_s\bbeta_s^\top W(t_m, \tau)\right\} \bx \otimes W(t_m, \tau) \right)
}
{\Delta \left(\exp\left\{\mathop{\sum}\limits_{s=0}^S x_s\bbeta_s^\top W(t_m, \tau)\right\} \right)
}, 
$$
and 
$$  
 \frac{
	\Delta \left(\exp\left\{\mathop{\sum}\limits_{s=0}^S x_s\bbeta_s^\top W(t_m, \tau) \right\}
	\mathop{\sum}\limits_{s=0}^S x_s (\beta_{s2}-\beta_{s1})\tau^{-1} I(t_m\leq \tau)  \right)
}
{
	\Delta  \left(\exp\left\{\mathop{\sum}\limits_{s=0}^S x_s\bbeta_s^\top W(t_m, \tau) \right\} \right)
}
$$ 
are finite. Thus, by the mean-value theorem, 
$\left|m_{\btheta_1}(\mathcal{Z}) - m_{\btheta_2}(\mathcal{Z})\right|\leq m(\mathcal{Z})\|\btheta_1-\btheta_2\|$ for every $\mathcal{Z}$, 
where 
\[
	m(\mathcal{Z})=
  \begin{bmatrix}
   \max \left| 
   	\mathop{\sum}\limits_{m=1}^M \mathop{\sum}\limits_{s=0}^S
   U_{im}
   \frac{ 
   	\Delta \left(\exp\left\{\mathop{\sum}\limits_{s=0}^S x_s\bbeta_s^\top W(t_m, \tau)\right\}\bx\otimes W(t_m, \tau) \right)
   }
   {\Delta \left(\exp\left\{\mathop{\sum}\limits_{s=0}^S x_s\bbeta_s^\top W(t_m, \tau)\right\} \right)
   }
   \right|  & \\
   \max\left| 
     \mathop{\sum}\limits_{m=1}^M \mathop{\sum}\limits_{s=0}^S
    U_{im}
    \frac{
    	\Delta \left(\exp\left\{\mathop{\sum}\limits_{s=0}^S x_s\bbeta_s^\top W(t_m, \tau) \right\}
    	\mathop{\sum}\limits_{s=0}^S x_s(\beta_{s2}-\beta_{s1})\tau^{-1} I(t_m\leq \tau)  \right)
    }
    {
    	\Delta  \left(\exp\left\{\mathop{\sum}\limits_{s=0}^S x_s\bbeta_s^\top W(t_m, \tau) \right\} \right)
    }
	\right| & 
  \end{bmatrix}
  <   \infty.
\]
Then $m_{\btheta}$ is Lipschitz continuous with respect to $\btheta$.  

Since $\Theta$ is compact and the uniqueness of the minimum $\btheta_0$ (by Conditions A1 and A2), along with that $\ell_{n}(\btheta)$ is continuous with respective to $\btheta$,  
then we can establish that  $\widehat{\btheta}\mathop{\longrightarrow}\limits^P \btheta$,  
by Theorem 2.1 of \citet{newey1994large}. 

(ii) For the asymptotic normality, we  apply Theorem 5.23 in \citet{van1998asymptotic} to establish the asymptotic normality of M-estimators in the case the criterion function is Lipschitz and its limit function admits a second order Taylor expansion.     
Indeed, the map $\btheta\longmapsto \mbox{E}m_{\btheta}=M_\tau(\btheta)$ is Lipschitz continuous with respect to $\btheta$ by (i). 
Furthermore,  the map $\btheta\longmapsto \mbox{E}m_{\btheta}=M_\tau(\btheta)$ also admits a second order Taylor expansion at $\btheta_0$, with a nonsingular symmetric Hessian matrix 
$A(\btheta_0)$.  
Thus, $\sqrt{n}(\widehat{\btheta}-\btheta_0)$ is asymptotically normal with mean zero and covariance matrix $A(\btheta_0)^{-1}\Sigma(\btheta_0) A(\btheta_0)^{-1}$. 

\section*{Appendix B.   Algorithm.}

\vspace{10pt}
\fbox{ 
	\begin{minipage}{\dimexpr\textwidth-2\fboxsep-2\fboxrule\relax} 
		\parbox{1\textwidth}{%
			\textbf{Algorithm~1}:
			\begin{center}  
				\begin{enumerate}
					\item
					Generate iid r.v.'s  $\{\xi_i: i=1,\ldots, n\}$  from  $N(0,1)$. 
					
					\item
					Simulate unrestricted and restricted score functions, respectively, 
					\[
					G(\tau)=\frac{1}{\sqrt{n}}\sum_{i=1}^n \xi_i \bigtriangledown_{\bbeta}\widehat{q}_i, 
					\quad \text{and} \quad 
					G_1(\tau)=\frac{1}{\sqrt{n}}\sum_{i=1}^n \xi_i \bigtriangledown_{\beta_1}\widetilde{q}_i, 
					\]
					where 
					\begin{align*}
					\begin{aligned}
					\bigtriangledown_{\bbeta}\widehat{q}_i &=
					\sum_{m=1}^M 
					U_{im}
					\frac{ 
						\Delta \left(\exp\left\{ \widehat{\bbeta}_0^\top W_i(t_m, \widehat{\tau}) \right\} W_i(t_m, \widehat{\tau}) \right)
					}
					{\Delta \left(\exp\left\{\widehat{\bbeta}_0^\top W_i(t_m, \widehat{\tau}) \right\} \right)
					},\\
					\bigtriangledown_{\beta_1}\widetilde{q}_i&=
					\sum_{m=1}^M 
					U_{im}
					\frac{ 
						\Delta \left(\exp\left\{\widetilde{\beta}_1 \log(t_m) \right\} \log(t_m)  \right)
					}
					{\Delta \left(\exp\left\{\widetilde{\beta}_1 h(t)\right\} \right)
					}. 
					\end{aligned}
					\end{align*}

					\item
					Simulate the test statistic 
					\[
					D= \sup_\tau
					\frac{1}{2}
					\left[
					G(\tau)^\top\widehat{I}^{-1}(\tau)G(\tau)-
					G_1(\tau)^\top\widehat{I}_1^{-1}(\tau)G_1(\tau),
					\right]
					\]
					where $\widehat{I}=n^{-1}\mathop{\sum}\limits_{i=1}^n \bigtriangledown_{\bbeta}\widehat{q}_i \bigtriangledown_{\bbeta}\widehat{q}_i^\top$ 
					and 
					$\widehat{I}_1=n^{-1}\mathop{\sum}\limits_{i=1}^n \bigtriangledown_{\beta_1}\widetilde{q}_i \bigtriangledown_{\beta_1}\widetilde{q}_i^\top$. 
					
					\item
					Repeat steps 1--3 for $NB$ times to obtain $D^{(1)},\cdots, D^{(NB)}$,  and calculate 
					$\widehat{p}= NB^{-1}\mathop{\sum}\limits_{j=1}^{NB} I(D^{(j)} >QLR)$. 
					
				\end{enumerate}
			\end{center}  
		} 
	\end{minipage}
}

\bibliographystyle{apalike}